\documentstyle[preprint,aps,eqsecnum,amssymb,amsfonts,verbatim]{revtex}
\tightenlines

\begin{document}

\title{\bf Rigged Hilbert Space Treatment of Continuous Spectrum}

\author{R.~de la Madrid}

\address{Institute for Scientific Interchange (ISI),
Villa Gualino, Viale Settimio Severo 65, I-10133, Torino, Italy \\
and \\
Departamento de F\'\i sica Te\'orica, Facultad de Ciencias, 47011 Valladolid, 
Spain \\
E-mail: \texttt{rafa@isiosf.isi.it}}
\author{A.~Bohm}

\address{Center for Particle Physics, The University of Texas at Austin,
Austin, Texas 78712-1081, USA}

\author{M.~Gadella}

\address{Departamento de F\'\i sica Te\'orica, Facultad de Ciencias, 47011 
Valladolid, Spain}

\date{March 16, 2002}

\maketitle

\begin{abstract}
\noindent The ability of the Rigged Hilbert Space formalism to deal with 
continuous spectrum is demonstrated within the example of the square barrier 
potential.  The non-square integrable solutions of the time-independent 
Schr\"odinger equation are used to define Dirac kets, which are (generalized) 
eigenvectors of the Hamiltonian.  These Dirac kets are antilinear functionals 
over the space of physical wave functions.  They are also basis vectors that 
expand any physical wave function in a Dirac basis vector expansion.  It is 
shown that an acceptable physical wave function must fulfill stronger 
conditions than just square integrability---the space of physical wave 
functions is not the whole Hilbert space but rather a dense subspace of the 
Hilbert space.  We construct the position and energy representations of the 
Rigged Hilbert Space generated by the square barrier potential Hamiltonian.  We
shall also construct the unitary operator that transforms from the position
into the energy representation.  We shall see that in the energy representation
the Dirac kets act as the antilinear Schwartz delta functional.  In 
constructing the Rigged Hilbert Space of the square barrier potential, we will 
find a systematic procedure to construct the Rigged Hilbert Space of a large 
class of spherically symmetric potentials.  The example of the square
barrier potential will also make apparent that the natural framework 
for the solutions of a Schr\"odinger operator with continuous
spectrum is the Rigged Hilbert Space rather than just the Hilbert space.  
\end{abstract}

\pacs{03.65.-w, 02.30.Hq}


\def\llra{\relbar\joinrel\longrightarrow}              
\def\mapright#1{\smash{\mathop{\llra}\limits_{#1}}}    
\def\mapup#1{\smash{\mathop{\llra}\limits^{#1}}}     
\def\mapupdown#1#2{\smash{\mathop{\llra}\limits^{#1}_{#2}}} 

\newpage

\def\thesection{\arabic{section}}
\section{Introduction}
\def\thesection{\arabic{section}}
\setcounter{equation}{0}
\label{sec:introduction}

In the late 1920s, Dirac introduced a new mathematical model of Quantum 
Mechanics based upon a uniquely smooth and elegant abstract algebra of 
linear operators defined on an infinite dimensional complex vector space 
equipped with an inner product norm~\cite{DIRAC}.  Dirac's abstract 
algebraic model of {\it bras} and {\it kets} (from the bracket notation 
for inner product) proved to be of great heuristic value 
in the ensuing years, especially in dealing with Hamiltonians whose spectrum
is continuous. 

The Hilbert space (HS) was the first mathematical idealization proposed for 
Quantum Mechanics~\cite{VON}.  However, as von Neumann explains in the 
introduction to his book~\cite{VON}, the HS theory and Dirac's formalism are 
two different things.  Although there were attempts to realize the 
Dirac formalism in Hilbert space, there were a number of serious problems 
resulting from the fact that the Hilbert space cannot allocate such things as 
bras, kets, the Dirac delta function or the Dirac basis vector expansion, 
all of which are essential in any physical formulation of Quantum Mechanics 
that deals with continuous spectrum.  Indeed in his textual 
presentation~\cite{DIRAC} Dirac himself states that ``the bra and ket 
vectors that we now use form a more general space than a Hilbert Space'' 
(see~\cite{DIRAC}, page 40).

In the late 1940s, L.~Schwartz gave a precise meaning to the Dirac delta 
function as a functional over a space of test functions 
(cf.~\cite{SCHWARTZ}).  This led to the development of a new branch of 
functional analysis, the theory of distributions~\cite{SCHWARTZ}.  About the 
same time, von Neumann published the theory of direct integral 
decompositions of a Hilbert space induced by a self-adjoint 
operator~\cite{VON2} (also valid for more general cases).  This spectral 
theory was closer to classical Fourier analysis and represented an 
improvement over the earlier von Neumann's spectral theory~\cite{VON}.

I.~Gelfand always thought that von Neumann's spectral theory was not the 
whole story of the theory of linear operators defined on infinite
dimensional vector spaces.  Prompted by the theory of distributions, he and
his school introduced the Rigged Hilbert Space (RHS).  Starting out with this
RHS and von Neumann's direct integral decomposition, they were able to prove
the Nuclear Spectral theorem (also called Gelfand-Maurin 
theorem)~\cite{GELFAND}.  This theorem justifies Dirac basis vector expansion.

One of the aspects of Dirac's formalism, the continuity of the elements of
the algebra of observables, was discussed in the early 1960s in 
Refs.~\cite{MEJLBO,KRISTENSEN}.  If two operators of the algebra of 
observables satisfy the canonical (Heisenberg) commutation relation, at least 
one of them cannot be continuous (bounded) with respect to the Hilbert space 
topology.  In Refs.~\cite{MEJLBO,KRISTENSEN}, it is shown that there are
subdomains of the Hilbert space that can be endowed with (locally convex) 
topologies that make those operators continuous;  the largest of those 
subdomains is the Schwartz space.

In the mid 1960s, some physicists~\cite{B60,ROBERTS,ANTOINE} independently 
realized that the RHS provides a rigorous mathematical rephrasing of
all the aspects of Dirac's formalism.  In particular, the Nuclear Spectral 
theorem restates Dirac basis vector expansion along with the Dirac bras and 
kets within a mathematical theory.  Later on, the RHS was used to accommodate 
resonance states (Gamow vectors) 
(cf.~\cite{BOHM81,BOHMGA,ANTONIOU93,BOLLINI96,BOHM97,ANTOINE98,DIS} and 
references therein).  Applications of the RHS formalism to Quantum Mechanics 
can now be found in some textbooks~\cite{BOHM,GALINDO}.

Although there are some explicit examples of RHS in the 
literature (see for instance~\cite{B70}), no example of the RHS generated by a 
Schr\"odinger Hamiltonian with continuous spectrum has been constructed 
yet.  Here we try to fill in this gap~\cite{DIS}.

The dynamical equation that governs the behavior of a quantum system at 
any time is the time-dependent Schr\"odinger equation,
\begin{equation}
      i\hbar \frac{\partial}{\partial t}\varphi (t) =H \varphi (t) \, ,
       \label{tdSequ}
\end{equation}
where $H$ denotes the Hamiltonian of the system and $\varphi (t)$ denotes 
the value of the wave function $\varphi$ at time $t$.  In order to solve
(\ref{tdSequ}), we associate to each energy $E$ in the spectrum ${\rm Sp}(H)$ 
of the Hamiltonian a ket $|E\rangle$ which is an eigenvector of $H$,
\begin{equation}
       H|E\rangle =E |E\rangle \, , \quad E\in {\rm Sp}(H) \, .
       \label{tiSequ}
\end{equation}  
These eigenkets form a complete basis system that expands any wave function 
$\varphi$ as
\begin{equation}
        \varphi =\int dE\, |E\rangle  \langle E|\varphi \rangle \equiv
       \int dE \, \varphi (E) |E\rangle \, .
       \label{DbveI}
\end{equation}
The time-dependent solution of Eq.~(\ref{tdSequ}) is obtained by 
Fourier-transforming the time-independent solution (\ref{DbveI}),
\begin{equation}
      \varphi (t)=\int dE\, e^{-iEt/\hbar}\, \varphi (E)  \, .
       \label{tiemdesodye} 
\end{equation}

If the spectrum of the Hamiltonian has a continuous part, and if the 
energy $E$ belongs to this continuous part of the spectrum, then the 
corresponding eigenket $|E\rangle$ that solves Eq.~(\ref{tiSequ}) is not 
square integrable, i.e., $|E\rangle$ is not an element of the Hilbert 
space.  Therefore, the eigenket $|E\rangle$ cannot represent 
an experimentally preparable physical state. 

We will show that the expansion (\ref{DbveI}) is only valid for those 
$\varphi$ that belong to a space of test functions 
${\mathbf \Phi} \subset {\cal H}$.  We will also show that the kets 
$|E\rangle$ can be understood mathematically as continuous
antilinear functionals over the space of test functions $\mathbf \Phi$, i.e.,
$|E\rangle \in \mathbf \Phi ^{\times}$.  According to the RHS mathematics, 
equation (\ref{tiSequ}) means that
\begin{equation}
      \langle H\varphi |E\rangle =E\langle \varphi |E\rangle \, , \quad 
      {\rm for \ every \ } \varphi \in \mathbf \Phi \, ,
      \label{geing}
\end{equation}
where $H$ is (the restriction to $\mathbf \Phi$ of) the 
self-adjoint Hamiltonian operator, which is a continuous operator on 
the linear topological space $\mathbf \Phi$.  For every such an operator, one 
defines the conjugate operator $H^{\times}$ on $\mathbf \Phi ^{\times}$ by
\begin{equation}
      \langle H\varphi |F\rangle =\langle \varphi |H^{\times}F\rangle \, ,
      \quad {\rm for \ all \ } \varphi \in {\mathbf \Phi} \, , \ F\in
       \mathbf \Phi ^{\times} \, .
       \label{def}
\end{equation}
The operator $H^{\times}$ is a uniquely defined extension of the Hilbert 
space adjoint
operator $H^{\dagger}$ (which for the case of a essentially self-adjoint 
operator coincides with the closure of $H$).  Using the definition (\ref{def}),
we write (\ref{geing}) formally as
\begin{equation}
       H^{\times}|E\rangle =E|E\rangle \, , \quad 
       |E\rangle \in \mathbf \Phi ^{\times} \, ,
       \label{timseqcors}
\end{equation}
which is understood as a functional equation over the space 
$\mathbf \Phi$.  The quantities $E$ and $|E\rangle$ are called generalized
eigenvalues and generalized eigenvectors, respectively.

The general statement of the Nuclear Spectral theorem just assures the 
existence of the generalized eigenvectors $|E\rangle$, but it does 
not provide a prescription to construct them.  In this paper, we construct
the generalized eigenvectors $|E\rangle$ of the square barrier Hamiltonian
along with the RHS.  We shall use the Sturm-Liouville theory
(Weyl theory)~\cite{DUNFORD} to find the RHS of the square barrier potential. 

By applying the Sturm-Liouville theory to the Schr\"odinger
equation of the square barrier potential, we will obtain a
domain ${\cal D}(H)$ on which the Hamiltonian is self-adjoint.  The 
Green functions, the spectrum, and the unitary transformation that 
diagonalizes our Hamiltonian will be also computed.  The diagonalization of the
Hamiltonian will allow us to obtain the energy (spectral) representation and
the direct integral decomposition of the Hilbert space induced by our
Hamiltonian.  We will see why this direct integral
decomposition is not enough for the purposes of Quantum Mechanics and why
the RHS is necessary.  Next, we will construct the space
$\mathbf \Phi$ and therewith the RHS of the square barrier potential,
\begin{equation}
       \mathbf \Phi \subset {\cal H} \subset \mathbf \Phi ^{\times} \, . 
        \label{GT}
\end{equation}
Dirac kets will be
accommodated as elements of $\mathbf \Phi ^{\times}$, and the Schwartz delta 
function will appear in the energy (spectral) representation of the triplet 
(\ref{GT}).  The Nuclear Spectral theorem will be proved, 
and it will be shown that this theorem is just a restatement of the heuristic 
Dirac basis vector expansion.

\def\thesection{\arabic{section}}
\section{Sturm-Liouville Theory Applied to the Square Barrier Potential}
\def\thesection{\arabic{section}}
\setcounter{equation}{0}
\label{sec:SLT}

\def\thesubsection{\thesection.\arabic{subsection}}
\subsection{Schr\"odinger Equation in the Position Representation}
\label{sec:SE}

In order to calculate the set of real generalized eigenvalues of the
square barrier Hamiltonian (the physical spectrum) and their corresponding 
generalized eigenvectors, we solve equation~(\ref{timseqcors}) in the position 
representation,  
\begin{equation}
      \langle \vec{x}|H^{\times}|E\rangle =E\langle \vec{x}|E\rangle \, .
      \label{Schrroenpso}
\end{equation}
The expression of the Hamiltonian in the position representation is
\begin{equation}
	\langle \vec{x}|H^{\times}|E\rangle =
	\left(\frac{-\hbar ^2}{2m} \Delta +V(\vec{x})\right)
	\langle \vec{x}|E\rangle  \, ,
	\label{delta}
\end{equation}
where $\Delta$ is the three-dimensional Laplacian and 
\begin{equation}
	V(\vec{x})=V(r)=\left\{ \begin{array}{ll}
                                0   &0<r<a  \\
                                V_0 &a<r<b  \\
                                0   &b<r<\infty 
                  \end{array} 
                 \right. 
	\label{potential}
\end{equation}
is the square barrier potential.  Writing Eqs.~(\ref{Schrroenpso}) and
(\ref{delta}) in spherical coordinates and restricting
ourselves to the case of zero angular momentum, we obtain the radial 
time-independent Schr\"odinger equation,
\begin{equation}
      \left( -\frac{\hbar ^2}{2m} \frac{d^2}{dr^2}+V(r) \right) \chi (r;E)=
      E\chi (r;E) \, .
      \label{rSe0}
\end{equation}
Thus our Hamiltonian in the radial representation is given by the 
differential operator
\begin{equation}
       h \equiv -\frac{\hbar ^2}{2m} \frac{d^2}{dr^2}+V(r) \, .
      \label{doh}
\end{equation}
Throughout this paper, the symbol $h$ will be used to denote the formal 
differential operator (\ref{doh}).  The formal differential operator
(\ref{doh}) is of the Sturm-Liouville type (cf.~\cite{DUNFORD}), and therefore
we are allowed to apply the Sturm-Liouville theory to our problem. 

Mathematically, all the information about the differential operator $h$ that 
is provided by the Sturm-Liouville theory (resolvent, spectrum, spectral 
representation,...) is obtained from the generalized eigenvalue equation
\begin{equation}
       h \chi (r;E)=
         \left( -\frac{\hbar ^2}{2m}\frac{d^2}{dr^2}+V(r) \right)
       \chi (r;E)=E\chi (r;E) \, , \quad E\in {\mathbb C} \, ,
       \label{info}
\end{equation}
subject to different boundary conditions.  From a physical point of view, 
Eq.~(\ref{info}) is the time-independent Schr\"odinger equation.   As 
mentioned in the introduction, the 
monoenergetic eigensolutions of (\ref{info}) are not in general square 
integrable, i.e., they are not in the Hilbert space.  Those 
monoenergetic eigensolutions will be associated to antilinear functionals 
$F_E \in \mathbf \Phi ^{\times}$ by
\begin{equation}
       F_E (\varphi ) \equiv \int_0^{\infty}dr\, 
       \overline{\varphi (r)} \chi (r;E) \, .
\end{equation}
These functionals are generalized eigenvectors of the Hamiltonian $H$,
\begin{equation}
      H^{\times}F_E=EF_E \, ,
\end{equation}
or more precisely,
\begin{equation}
      \langle \varphi |H^{\times}|F_E \rangle = \langle H\varphi |F_E \rangle
      =E\langle \varphi |F_E\rangle \, , \quad \forall \varphi \in 
      \mathbf \Phi \, .
\end{equation}

\def\thesubsection{\thesection.\arabic{subsection}}
\subsection{Self-Adjoint Extension}
\label{sec:SAE}

Our first objective will be to define a linear operator on
a Hilbert space corresponding to the formal differential operator
$h$ and investigate its self-adjoint extensions.  Among all the possibilities,
we shall choose the self-adjoint extension that fits spherically 
symmetric potentials.  Later sections will deal with the spectral properties 
of this self-adjoint extension and with the RHS induced by it. 

The Hilbert space that is in the RHS of the square barrier potential is 
realized by the space $L^2([0,\infty ), dr)$ of square integrable functions 
$f(r)$ defined on the interval $[0,\infty )$.  In this section, we find a 
subdomain ${\cal D}(H)$ of this Hilbert space on which the 
differential operator $h$ is self-adjoint.  This domain must be a proper 
dense linear subspace of $L^2([0,\infty ), dr)$.  The action of $h$ must be 
well-defined on ${\cal D}(H)$, and this action must remain 
in $L^2([0,\infty ), dr)$.  We need also a boundary condition that assures 
the self-adjointness of the Hamiltonian.  Among 
all the possible boundary conditions that provide a self-adjoint extension
(see Appendix~\ref{sec:A1}), we choose $f(0)=0$.  These requirements can be 
written as
\begin{mathletters}
      \label{bcthdpspr}
\begin{eqnarray}
      f(r) \in L^2([0,\infty ), dr)  \, , \label{HSc} \\
      hf(r) \in L^2([0,\infty ), dr) \, , \label{reduce} \\
       f(r) \in AC^2 [0,\infty ) \, ,  \label{ACc} \\
       f(0)=0 \, , \label{sac}  
\end{eqnarray}
\end{mathletters}
where $AC^2[0,\infty)$ denotes the space of functions whose derivative 
is absolutely continuous (see Appendix~\ref{sec:A1}).  Condition (\ref{HSc}) 
just means that the wave functions are square 
normalizable.  Condition (\ref{reduce}) assures that the action of $h$ on 
any $f(r)\in {\cal D}(H)$ is square integrable.  Condition (\ref{ACc}) is the 
weakest condition sufficient for the second derivative of $f(r)$ to be 
well-defined.  In our example, this condition implies that $f(r)$ and 
$f'(r)$ are continuous at $r=a$ and at $r=b$.  Condition (\ref{sac}) 
selects the self-adjoint extension needed in physics.

The reason why we choose (\ref{sac}) is the following:  in 
physics  \cite{DIRAC,NEWTON,COHEN,TAYLOR}, the set of boundary conditions
imposed on the Schr\"odinger equation (\ref{info}) always includes
\begin{mathletters}
      \label{pshybouc}
\begin{eqnarray}
      && \chi (0;E)=0 \, , \label{prc} \\
      && \chi (r;E), {\rm \ and \ }  \chi '(r;E) 
      {\rm \ are \ continuous \ at \ } r=a 
       {\rm \ and \ at \ } r=b \, .  \label{pcc}      
\end{eqnarray}
\end{mathletters}
Condition (\ref{pcc}) is implied by (\ref{ACc}), so we just need to
recover (\ref{prc}).  This is why we impose (\ref{sac}).

The set of conditions (\ref{bcthdpspr}) leads to the domain
\begin{equation}
       {\cal D}(H) =\{ f(r)\, | \ f(r), hf(r)\in L^2([0,\infty ), dr), \,
                   f(r) \in AC^2[0,\infty ), \, f(0)=0 \} \, .
     \label{domain}
\end{equation}
In choosing (\ref{domain}) as the domain of our formal differential
operator $h$, we define a linear operator $H$ by
\begin{equation}
      Hf(r) :=hf(r)= 
      \left( -\frac{\hbar ^2}{2m}\frac{d^2}{dr^2} +V(r) \right)f(r)
       \, , \quad f(r) \in {\cal D}(H) \, .
      \label{operator}
\end{equation}

\def\thesubsection{\thesection.\arabic{subsection}}
\subsection{Resolvent and Green Functions}
\label{sec:RaGF}

The Green function is the kernel of integration needed to write the 
resolvent of $H$ as an integral operator,
\begin{equation}
      \left( E-H \right) ^{-1}f(r)=\int_0^{\infty}G(r,s;E)f(s)\, ds \, .
\end{equation}
The procedure to compute the Green function of our operator 
(\ref{operator}) is explained in~\cite{DUNFORD} (see also~\cite{CSF}).  For 
the sake of completeness,
we include in Appendix~\ref{sec:A2} the theorem that is used 
to calculate $G(r,s;E)$.  

The expression of the Green function will be given in terms of eigenfunctions 
of the differential operator $h$ subject to different boundary conditions
(see Theorem~1 of Appendix~\ref{sec:A2}).  We shall consider three regions of 
the complex plane and compute the Green
function for each region separately.  In all our calculations, we will 
use the following branch of the square root function:
\begin{equation}
      \sqrt{\cdot}:\{ E\in {\mathbb C} \, | \  -\pi <{\rm arg}(E)\leq \pi \} 
   \longmapsto \{E\in {\mathbb C} \, | \  -\pi/2 <{\rm arg}(E)\leq \pi/2 \} 
     \, . 
\end{equation}

\vskip0.4cm

\centerline{\bf Region $\Re (E)<0$, $\Im (E)\neq 0$}

\vskip0.2cm

For $\Re (E)<0$, $\Im (E)\neq 0$, the Green function (see Theorem~1 of
Appendix~\ref{sec:A2}) is given by
\begin{equation}
       G(r,s;E)=\left\{ \begin{array}{ll}
      -\frac{2m/\hbar ^2}{\sqrt{-2m/\hbar ^2 \, E}} \,
      \frac{\widetilde{\chi}(r;E) \, \widetilde{\Theta} (s;E)}
       {2 \widetilde{{\cal J}}_3(E)}
      \quad  &r<s \\ [2ex]
     -\frac{2m/\hbar ^2}{\sqrt{-2m/\hbar ^2 \, E}}\, 
   \frac{\widetilde{\chi}(s;E)\,  \widetilde{\Theta} (r;E)}
          {2 \widetilde{{\cal J}}_3(E)}
     \quad &r>s 
        \end{array} 
      \right. 
      \quad \Re (E)<0 \, , \ \Im (E)\neq 0 \, .
	\label{-}
\end{equation}
The eigenfunction $\widetilde{\chi}(r;E)$ satisfies the Schr\"odinger 
equation (\ref{info}) and the boundary conditions
\begin{mathletters}
      \label{eigsbo0co}
\begin{eqnarray}
       && \widetilde{\chi} (0;E)=0 \, , \label{bca01} \\
       && \widetilde{\chi} (r;E)\in AC^2([0,\infty )) \, ,\\
       && \widetilde{\chi}(r;E)
           {\rm \ is \ square \ integrable \ at \ } 0 \, .
       \label{sbca03}  
\end{eqnarray}
\end{mathletters}
The boundary conditions (\ref{eigsbo0co}) can be written as
\begin{mathletters}
      \label{bctchiphis}
\begin{eqnarray}
      \widetilde{\chi} (0;E)&=&0 \, ,  \\
      \widetilde{\chi} (a-0;E)&=& \widetilde{\chi} (a+0;E) \, ,
        \label{bocchiata1}\\
      \widetilde{\chi}' (a-0;E)&=& \widetilde{\chi}' (a+0;E) \, ,\\
      \widetilde{\chi} (b-0;E)&=& \widetilde{\chi} (b+0;E) \, ,\\
      \widetilde{\chi}' (b-0;E)&=& \widetilde{\chi}' (b+0;E) \, ,
         \label{bocchiata4}\\
      \widetilde{\chi}(r;E) && \!\!\!\!\!
           \mbox{is square integrable at}\  0 \, , 
\end{eqnarray}
\end{mathletters}
and lead to
\begin{equation}
      \widetilde{\chi}(r;E) =\left\{ \begin{array}{lll}
               e^{\sqrt{- \frac{2m}{\hbar ^2}E}r}-
                e^{-\sqrt{- \frac{2m}{\hbar ^2}E}r} \quad &0<r<a  \\
            \widetilde{{\cal J}}_1(E)e^{\sqrt{- \frac{2m}{\hbar ^2}(E-V_0)}r}
            +\widetilde{{\cal J}}_2(E)e^{-\sqrt{- \frac{2m}{\hbar ^2}(E-V_0)}r}
                 \quad  &a<r<b \\
            \widetilde{{\cal J}}_3(E)e^{\sqrt{- \frac{2m}{\hbar ^2}E}r}
            +\widetilde{{\cal J}}_4(E)e^{-\sqrt{- \frac{2m}{\hbar ^2}E}r}
                 \quad  &b<r<\infty \, .
               \end{array} 
                 \right. 
       \label{tildechifunction}
\end{equation}
The functions $\widetilde{{\cal J}}_1-\widetilde{{\cal J}}_4$ are such
that $\widetilde{\chi}(r;E)$ satisfies the boundary conditions 
(\ref{bctchiphis}), and their expressions are given by
Eq.~(\ref{tildejfunc}) of Appendix~\ref{sec:A2}.

The eigenfunction $\widetilde{\Theta}(r;E)$ satisfies the Schr\"odinger
equation (\ref{info}) and the boundary conditions
\begin{mathletters}
       \label{bcainfty1}
\begin{eqnarray}
       &&\widetilde{\Theta}(r;E)\in AC^2([0,\infty )) \, , \\
       &&\widetilde{\Theta} (r;E) \ 
         {\rm is \ square \ integrable \ at \ } \infty \, .
\end{eqnarray}
\end{mathletters}
The boundary conditions (\ref{bcainfty1}) can be written as
\begin{mathletters}
        \label{thetispsi}
\begin{eqnarray}
      \widetilde{\Theta} (a-0;E)&=& \widetilde{\Theta} (a+0;E) \, , 
       \label{condtThea}\\
      \widetilde{\Theta}' (a-0;E)&=& \widetilde{\Theta}' (a+0;E) \, ,\\
      \widetilde{\Theta} (b-0;E)&=& \widetilde{\Theta} (b+0;E) \, ,\\
      \widetilde{\Theta}' (b-0;E)&=& \widetilde{\Theta}' (b+0;E) \, ,
        \label{condtTheb}\\
      \widetilde{\Theta}(r;E) && \!\!\!\!\!
           \mbox{is square integrable at} \ \infty \, , 
\end{eqnarray}
\end{mathletters}
and lead to
\begin{equation}
      \widetilde{\Theta}(r;E)=\left\{ \begin{array}{lll}
         \widetilde{{\cal A}}_1(E)e^{\sqrt{- \frac{2m}{\hbar ^2}E}r}
          +\widetilde{{\cal A}}_2(E)e^{-\sqrt{- \frac{2m}{\hbar ^2}E}r} 
           \quad &0<r<a  \\
         \widetilde{{\cal A}}_3(E)e^{\sqrt{- \frac{2m}{\hbar ^2}(E-V_0)}r}
         +\widetilde{{\cal A}}_4(E)e^{-\sqrt{- \frac{2m}{\hbar ^2}(E-V_0)}r}
                 \quad  &a<r<b \\
               e^{-\sqrt{- \frac{2m}{\hbar ^2}E}r}
                 \quad  &b<r<\infty \, . 
               \end{array} 
                 \right. 
      \label{tildethetfunc}
\end{equation}
The functions $\widetilde{{\cal A}}_1-\widetilde{{\cal A}}_4$ are such that 
$\widetilde{\Theta}(r;E)$ satisfies the boundary conditions 
(\ref{thetispsi}), and their expressions are given by
Eq.~(\ref{tildeAfunc}) of Appendix~\ref{sec:A2}. 

\vskip0.5cm

\centerline{\bf Region $\Re (E)>0$, $\Im (E)> 0$}

\vskip0.2cm

When $\Re (E)>0$, $\Im (E)>0$, the expression of the Green function is
\begin{equation}
       G(r,s;E)=\left\{ \begin{array}{ll}
       \frac{2m/\hbar ^2}{\sqrt{2m/\hbar ^2 \, E}} \,
        \frac{\chi (r;E) \, \Theta _+ (s;E)}{2i {\cal J}_4(E)}
         &r<s  \\ [2ex]
       \frac{2m/\hbar ^2}{\sqrt{2m/\hbar ^2 \, E}} \,
      \frac{\chi (s;E) \, \Theta _+ (r;E)}{2i {\cal J}_4(E)}
               &r>s 
         \end{array} 
                 \right. 
       \quad \Re (E)>0, \ \Im (E)>0 \, . 
	\label{++}
\end{equation}
The eigenfunction $\chi (r;E)$ satisfies the Schr\"odinger equation 
(\ref{info}) and the boundary conditions (\ref{eigsbo0co}),
\begin{equation}
      \chi (r;E)=\left\{ \begin{array}{lll}
               \sin (\sqrt{\frac{2m}{\hbar ^2}E}r) \quad &0<r<a  \\
               {\cal J}_1(E)e^{i \sqrt{\frac{2m}{\hbar ^2}(E-V_0)}r}
                +{\cal J}_2(E)e^{-i\sqrt{\frac{2m}{\hbar ^2}(E-V_0)}r}
                 \quad  &a<r<b \\
               {\cal J}_3(E) e^{i\sqrt{\frac{2m}{\hbar ^2}E}r}
                +{\cal J}_4(E)e^{-i\sqrt{\frac{2m}{\hbar ^2}E}r}
                 \quad  &b<r<\infty \, .
               \end{array} 
                 \right. 
             \label{chi}
\end{equation}
The functions ${\cal J}_1-{\cal J}_4$ are determined by the boundary 
conditions (\ref{bctchiphis}), and their expressions are listed in
Eq.~(\ref{Jfunction}) of Appendix~\ref{sec:A2}.  

The eigenfunction $\Theta _+(r;E)$ satisfies the Schr\"odinger equation
(\ref{info}) and the boundary conditions (\ref{bcainfty1}),
\begin{equation}
      \Theta _+(r;E)=\left\{ \begin{array}{lll}
               {\cal A}^+_1(E)e^{i\sqrt{\frac{2m}{\hbar ^2}E}r}
               +{\cal A}^+_2(E) e^{-i\sqrt{\frac{2m}{\hbar ^2}E}r} 
                \quad &0<r<a  \\
               {\cal A}^+_3(E)e^{i\sqrt{\frac{2m}{\hbar ^2}(E-V_0)}r}
                +{\cal A}^+_4(E)e^{-i\sqrt{\frac{2m}{\hbar ^2}(E-V_0)}r}
                 \quad  &a<r<b \\
               e^{i\sqrt{\frac{2m}{\hbar ^2}E}r}
                 \quad  &b<r<\infty \, . 
               \end{array} 
                 \right. 
            \label{theta+fun}
\end{equation}
The functions ${\cal A}^+_1 - {\cal A}^+_4$ are determined by the boundary
conditions (\ref{thetispsi}), and their expressions are listed in
Eq.~(\ref{A+functions}) of Appendix~\ref{sec:A2}. 

\vskip0.3cm

\centerline{\bf Region $\Re (E)>0$, $\Im (E)< 0$}

\vskip0.2cm

In the region $\Re (E)>0$, $\Im (E)<0$ the Green function reads
\begin{equation}
       G(r,s;E)=\left\{ \begin{array}{ll}
       -\frac{2m/\hbar ^2}{\sqrt{2m/\hbar ^2 \, E}} \,
        \frac{\chi (r;E) \, \Theta _- (s;E)}{2i {\cal J}_3(E)}
         &r<s  \\ [2ex]
       -\frac{2m/\hbar ^2}{\sqrt{2m/\hbar ^2 \, E}} \,
      \frac{\chi (s;E) \, \Theta _- (r;E)}{2i {\cal J}_3(E)}
               &r>s 
         \end{array} 
                 \right. 
       \quad \Re (E)>0, \ \Im (E)<0 \, . 
	\label{+-}
\end{equation}
The eigenfunction $\chi (r;E)$ is given by (\ref{chi}).  The 
eigenfunction $\Theta _-(r;E)$ satisfies the Schr\"odinger equation 
(\ref{info}) and the boundary conditions (\ref{bcainfty1}),
\begin{equation}
      \Theta _-(r;E)=\left\{ \begin{array}{lll}
               {\cal A}_1^-(E)e^{i\sqrt{\frac{2m}{\hbar ^2}E}r}
               +{\cal A}_2^-(E) e^{-i\sqrt{\frac{2m}{\hbar ^2}E}r} 
                \quad &0<r<a  \\
               {\cal A}_3^-(E)e^{i\sqrt{\frac{2m}{\hbar ^2}(E-V_0)}r}
                +{\cal A}_4^-(E)e^{-i\sqrt{\frac{2m}{\hbar ^2}(E-V_0)}r}
                 \quad  &a<r<b \\
               e^{-i\sqrt{\frac{2m}{\hbar ^2}E}r}
                 \quad  &b<r<\infty \, . 
               \end{array} 
                 \right. 
        \label{thetafun-}
\end{equation}
The functions ${\cal A}^-_1-{\cal A}^-_4$ are such that $\Theta _-(r;E)$ and 
its derivative are continuous at $r=a$ and at $r=b$.  Their expressions are 
listed in Eq.~(\ref{A-functions}) of Appendix~\ref{sec:A2}.

\def\thesubsection{\thesection.\arabic{subsection}}
\subsection{Diagonalization of $H$ and Eigenfunction Expansion}
\label{sec:SRD}

In the present section, we diagonalize the Hamiltonian and construct the 
eigenfunction expansion generated by the eigenfunctions
of the differential operator $h$.  In order to do so, we compute the 
spectrum of $H$ and then construct a unitary operator $U$ that transforms 
from the position representation into the energy representation.  We
will see that the spectrum of $H$ is the positive real line $[0,\infty )$.  In
the energy representation, $H$ will act as the multiplication operator, the 
Hilbert space will be realized by $L^2([0,\infty ),dE)$ and the domain of the 
Hamiltonian will be realized by the maximal domain on which the 
multiplication operator is well-defined.  On our way, we shall take advantage
of some theorems of the Sturm-Liouville theory that are proved in 
Ref.~\cite{DUNFORD}.  For the sake of completeness, we include those theorems 
in Appendix~\ref{sec:A3}.

\def\thesubsubsection{\thesubsection.\arabic{subsubsection}}
\subsubsection{\bf Spectrum of $H$}

We first compute the spectrum ${\rm Sp}(H)$ of the operator $H$ by applying 
Theorem~4 of Appendix~\ref{sec:A3} (see also \cite{DUNFORD}).  Since $H$ is 
self-adjoint, its spectrum is real.  The spectrum is the subset of 
the real line on which the Green function fails to be analytic.  This 
non-analyticity of $G(r,s;E)$ will be built into the functions $\theta _{ij}^{\pm}(E)$ that 
appear in Theorem~4 of Appendix~\ref{sec:A3}. 

From the expression of the Green function computed above, it is clear  that
the subsets $(-\infty ,0)$ and $(0,\infty )$ should be studied separately.  We
will denote either of these subsets by $\Lambda$.

\vskip0.5cm

\centerline{{\bf Subset } $\Lambda =(-\infty ,0)$}

\vskip0.3cm  

We first take $\Lambda$ from Theorem~4 of Appendix~\ref{sec:A3} to be 
$(-\infty ,0)$.  We choose a basis for the space of solutions of the 
equation $h\sigma =E\sigma$ that is continuous on $(0,\infty )\times \Lambda$ 
and analytically dependent on $E$ as
\begin{mathletters}
\begin{eqnarray}
      &&\sigma _1(r;E)=\left\{ \begin{array}{lll}
               \widetilde{\cal B}_1(E)e^{\sqrt{-\frac{2m}{\hbar ^2}E}r}
                +\widetilde{\cal B}_2(E)e^{-\sqrt{-\frac{2m}{\hbar ^2}E}r} 
                \quad &0<r<a  \\
               \widetilde{\cal B}_3(E)e^{\sqrt{-\frac{2m}{\hbar ^2}(E-V_0)}r}
                +\widetilde{\cal B}_4(E)
                 e^{-\sqrt{-\frac{2m}{\hbar ^2}(E-V_0)}r}
                 \quad  &a<r<b \\
               e^{\sqrt{-\frac{2m}{\hbar ^2}E}r}
                 \quad  &b<r<\infty   \, ,
               \end{array} 
                 \right. \qquad  \label{tildesigma1} \\  [2ex]
      &&\sigma _2(r;E)=\widetilde{\Theta}(r;E) \, .
\end{eqnarray}
\end{mathletters}
The functions $\widetilde{\cal B}_1-\widetilde{\cal B}_4$ are such that
$\sigma _1(r;E)$ and its derivative are continuous at $r=a$ and
at $r=b$.  Their expressions are listed in Eq.~(\ref{tildeBfunctions}) of
Appendix~\ref{sec:A3}.  The function $\widetilde{\Theta}(r;E)$ is 
given by Eq.~(\ref{tildethetfunc}).  

Obviously,
\begin{equation}
      \widetilde{\chi}(r;E)=\widetilde{{\cal J}}_3(E)\sigma _1(r;E)+
      \widetilde{{\cal J}}_4(E) \sigma _2(r;E) \, ,
\end{equation}
which along with Eq.~(\ref{-}) leads to
\begin{eqnarray}
      &&G(r,s;E)= 
      -\frac{2m/\hbar ^2}{\sqrt{-2m/\hbar ^2 \, E}} \,
       \frac{1}{2}\, \left[ \sigma _1(r;E)
       +\frac{\widetilde{{\cal J}}_4(E)}{\widetilde{{\cal J}}_3(E)} \, 
       \sigma _2(r;E)\right] \sigma _2(s;E) \, , 
       \nonumber \\
        &&\quad \hskip7cm r<s \, , \  \Re (E)<0 \, , \Im (E) \neq 0 \, . 
        \label{dis-parosof}
\end{eqnarray}
Since
\begin{equation}
      \overline{\sigma _2(s;\overline{E})}=
      \sigma _2(s;E) \, ,
\end{equation}
we can write Eq.~(\ref{dis-parosof}) as
\begin{eqnarray}
       &&G(r,s;E)= 
      -\frac{2m/\hbar ^2}{\sqrt{-2m/\hbar ^2 \, E}} \,
       \frac{1}{2}\,  \left[ \sigma _1(r;E)
       \overline{\sigma _2(s;\overline{E})}
       +\frac{\widetilde{{\cal J}}_4(E)}{\widetilde{{\cal J}}_3(E)} \, 
       \sigma _2(r;E)
       \overline{\sigma _2(s;\overline{E})} \right] , \nonumber \\
       &&\qquad \hskip7cm
        r<s \, , \  \Re (E)<0 \, , \Im (E) \neq 0 \, .
        \label{finag-retoco} 
\end{eqnarray}
On the other hand, by Theorem~4 in Appendix~\ref{sec:A3} we have
\begin{equation}
       G(r,s;E)=\sum_{i,j=1}^{2}
       \theta _{ij}^- (E)\sigma _i(r;E)
        \overline{\sigma _j(s;\overline{E})}\, ,
       \qquad r<s\, .
       \label{gf--}
\end{equation}
By comparing Eqs.~(\ref{finag-retoco}) and (\ref{gf--}) we see that
\begin{equation}
      \theta _{ij}^-(E)= \left(  \begin{array}{cl}
        0 & -\frac{2m/\hbar ^2}{\sqrt{-2m/\hbar ^2 \, E}} 
       \frac{1}{2}  \\
        0 &
      -\frac{2m/\hbar ^2}{\sqrt{-2m/\hbar ^2 \, E}} 
       \frac{1}{2}  
       \frac{\widetilde{\cal J}_4(E)}{\widetilde{\cal J}_3(E)} 
                                \end{array}
        \right) \, , \quad \Re (E)<0 \, , \  \Im (E) \neq 0 \, .
\end{equation}
The functions $\theta _{ij}^-(E)$ are analytic in a neighborhood of 
$\Lambda =(-\infty ,0)$.  Therefore, the interval $(-\infty ,0)$ is in the 
resolvent set ${\rm Re}(H)$ of the operator $H$.

\vskip0.5cm

\centerline{{\bf Subset} $\Lambda =(0, \infty )$}

\vskip0.3cm  

Now we study the case $\Lambda =(0,\infty )$.  In order to be able to 
apply Theorem 4 of Appendix~\ref{sec:A3}, we choose the following basis
for the space of solutions of $h\sigma =E\sigma$ that is continuous
on $(0,\infty )\times \Lambda$ and analytically dependent on $E$:  
\begin{mathletters}
      \label{basisin0infisi}
\begin{eqnarray}
      &&\sigma _1(r;E)=\chi (r;E)\, ,
      \label{sigma1=sci} \\ [2ex]
      &&\sigma _2(r;E)=\left\{ \begin{array}{lll}
               \cos (\sqrt{\frac{2m}{\hbar ^2}E}r) \quad &0<r<a  \\
               {\cal C}_1(E)e^{i \sqrt{\frac{2m}{\hbar ^2}(E-V_0)}r}
                +{\cal C}_2(E)e^{-i\sqrt{\frac{2m}{\hbar ^2}(E-V_0)}r}
                 \quad  &a<r<b \\
               {\cal C}_3(E)e^{i\sqrt{\frac{2m}{\hbar ^2}E}r}
                 +{\cal C}_4(E)e^{-i\sqrt{\frac{2m}{\hbar ^2}E}r}
                 \quad  &b<r<\infty \, . 
               \end{array} 
                 \right. \quad
       \label{sigam2cos} 
\end{eqnarray}
\end{mathletters}
The functions ${\cal C}_1-{\cal C}_4$, whose expressions are given by
Eq.~(\ref{Cfunctions}) of Appendix~\ref{sec:A3}, are such that $\sigma _2$ 
and its derivative are continuous at $r=a$ and at $r=b$.  The eigenfunction
$\chi (r;E)$ is given by Eq.~(\ref{chi}).  

Eqs.~(\ref{theta+fun}), (\ref{thetafun-}) and (\ref{basisin0infisi}) lead to  
\begin{equation}
      \Theta _+(r;E)=-\frac{{\cal C}_4(E)}{W(E)} \sigma _1(r;E)+
      \frac{{\cal J}_4(E)}{W(E)}\sigma _2(r;E) 
      \label{Thet+inosigm}
\end{equation}
and to
\begin{equation}
      \Theta _-(r;E)=\frac{{\cal C}_3(E)}{W(E)} \sigma _1(r;E)-
      \frac{{\cal J}_3(E)}{W(E)}\sigma _2(r;E) \, ,
      \label{Thet-inosigm}
\end{equation}
where
\begin{equation}
       W(E)={\cal J}_4(E){\cal C}_3(E)-{\cal J}_3(E){\cal C}_4(E) \, .
\end{equation}
By substituting Eq.~(\ref{Thet+inosigm}) into Eq.~(\ref{++}) we get to
\begin{eqnarray}
      &&G(r,s;E)=
       \frac{2m/\hbar ^2}{\sqrt{2m/\hbar ^2 \, E}} \,
       \frac{1}{2i{\cal J}_4(E)}\,
       \left[ -\frac{ {\cal C}_4(E)}{W(E)}
       \sigma _1(r;E)+\frac{ {\cal J}_4(E)}{W(E)}\sigma _2(r;E)\right] 
       \sigma _1 (s;E) \, ,  \nonumber \\
      &&\qquad \hskip7cm  \Re (E)>0, \Im (E)>0\, , \, r>s \, .
        \label{G++tofpuon}
\end{eqnarray}
By substituting Eq.~(\ref{Thet-inosigm}) into
Eq.~(\ref{+-}) we get to
\begin{eqnarray}
      &&G(r,s;E)=
      -\frac{2m/\hbar ^2}{\sqrt{2m/\hbar ^2 \, E}}   \,
       \frac{1}{2i{\cal J}_3(E)}\,
       \left[ \frac{ {\cal C}_3(E)}{W(E)}
       \sigma _1(r;E)-\frac{{\cal J}_3(E)}{W(E)}\sigma _2(r;E)\right]      
       \sigma _1 (s;E) \, , \nonumber \\
      &&\qquad \hskip7cm \Re (E)>0, \Im (E)<0\, , \, r>s \, ,
      \label{G+-tofpuon}
\end{eqnarray} 
Since
\begin{equation}
      \overline{\sigma _1(s;\overline{E})}=\sigma _1(s;E) \, ,
\end{equation}
Eq.~(\ref{G++tofpuon}) leads to
\begin{eqnarray}
      &&G(r,s;E)=
       \frac{2m/\hbar ^2}{\sqrt{2m/\hbar ^2 \, E}} \,
       \frac{1}{2i{\cal J}_4(E)}\,
       \left[ -\frac{ {\cal C}_4(E)}{W(E)}
       \sigma _1(r;E)\overline{\sigma _1(s;\overline{E})}   
      +\frac{ {\cal J}_4(E)}{W(E)}\sigma _2(r;E)
       \overline{\sigma _1(s;\overline{E})}\right] \nonumber \\      
      &&\qquad \hskip8cm \Re (E)>0, \Im (E)>0\, , \, r>s \, ,
       \label{redaot++}
\end{eqnarray}
and Eq.~(\ref{G+-tofpuon}) leads to 
\begin{eqnarray}
      &&G(r,s;E)=
      -\frac{2m/\hbar ^2}{\sqrt{2m/\hbar ^2 \, E}} \,
       \frac{1}{2i{\cal J}_3(E)}\,
       \left[ \frac{ {\cal C}_3(E)}{W(E)}
       \sigma _1(r;E)\overline{\sigma _1(s;\overline{E})}
       -\frac{{\cal J}_3(E)}{W(E)}\sigma _2(r;E)
       \overline{\sigma _1(s;\overline{E})}\right] \nonumber \\
       &&\qquad \hskip8cm \Re (E)>0, \Im (E)<0\, , \, r>s \, ,
       \label{redaot+-}
\end{eqnarray}
The expression of the resolvent in terms of the basis $\sigma _1,\sigma _2$
can be written as (see Theorem~4 in Appendix~\ref{sec:A3})
\begin{equation}
       G(r,s;E)=\sum_{i,j=1}^{2}
       \theta _{ij}^+ (E)\sigma _i(r;E)\overline{\sigma _j(s;\overline{E})}\, ,
       \quad r>s \, .
        \label{GF++}
\end{equation}
By comparing (\ref{GF++}) to (\ref{redaot++}) we get to
\begin{equation}
      \theta _{ij}^+(E)= \left(  \begin{array}{lc}
      \frac{2m/\hbar ^2}{\sqrt{2m/\hbar ^2 \, E}}
      \frac{1}{2i} \frac{- {\cal C}_4(E)}{ {\cal J}_4(E)W(E)}
       & 0 \\
      \frac{2m/\hbar ^2}{\sqrt{2m/\hbar ^2 \, E}}
      \frac{1}{2i} \frac{1}{W(E)}
       & 0
                                \end{array}
        \right) 
       \, , \quad  \Re (E)>0 \, , \  \Im (E)>0 \, ,
       \label{theta++}
\end{equation}
By comparing (\ref{GF++}) to (\ref{redaot+-}) we get to
\begin{equation}
      \theta _{ij}^+(E)= \left(  \begin{array}{lc}
      -\frac{2m/\hbar ^2}{\sqrt{2m/\hbar ^2 \, E}}
      \frac{1}{2i} \frac{{\cal C}_3(E)}{ {\cal J}_3(E)W(E)}
       & 0 \\
      \frac{2m/\hbar ^2}{\sqrt{2m/\hbar ^2 \, E}}
      \frac{1}{2i} \frac{1}{W(E)}
       & 0
                                \end{array}
        \right) 
       \, , \quad  \Re (E)>0 \, , \  \Im (E)<0 \, ,
      \label{theta+-}
\end{equation}
From Eqs.~(\ref{theta++}) and (\ref{theta+-}) we can see that the measures 
$\rho _{12}$, $\rho _{21}$ and $\rho _{22}$ in Theorem~4 of 
Appendix~\ref{sec:A3} are zero and that the measure $\rho _{11}$ is given by
\begin{eqnarray}
       \rho _{11}((E_1,E_2))&=&\lim _{\delta \to 0} \lim _{\epsilon \to 0+}
      \frac{1}{2\pi i} \int_{E_1+\delta}^{E_2-\delta}
      \left[ \theta _{11}^+ (E-i\epsilon ) -\theta _{11}^+ (E+i\epsilon )
      \right] dE \nonumber \\
      &=&\int_{E_1}^{E_2}   \frac{1}{4\pi}\, 
      \frac{2m/\hbar ^2}{\sqrt{2m/\hbar ^2 \, E}}
      \, \frac{1}{ {\cal J}_3(E) {\cal J}_4(E)}\, dE \, ,
\end{eqnarray}
which leads to
\begin{equation}
      \rho (E)\equiv \rho _{11}(E)=
      \frac{1}{4\pi}\, 
      \frac{2m/\hbar ^2}{\sqrt{2m/\hbar ^2 \, E}}
      \, \frac{1}{ |{\cal J}_4(E)|^2}\, , \quad E\in (0,\infty )  \, .
\end{equation} 
The function $\theta _{11}^+(E)$ has a branch cut along $(0,\infty)$, and 
therefore $(0,\infty )$ is included in ${\rm Sp}(H)$.  Since ${\rm Sp}(H)$ is 
a closed set, ${\rm Sp}(H)=[0,\infty )$.  Thus the resolvent 
set of $H$ is ${\rm Re}(H)={\mathbb C} -[0,\infty )$.

\subsubsection{\bf Diagonalization and Eigenfunction Expansion}

We are now in a position to diagonalize the Hamiltonian.  By Theorem~2 of 
Appendix~\ref{sec:A3}, there is a unitary map $\widetilde U$ defined by
\begin{eqnarray}
     \widetilde{U}:L^2([0,\infty ),dr) 
      &\longmapsto & L^2( (0,\infty ),\rho (E)dE) \nonumber \\
       f(r)& \longmapsto & 
         \widetilde{f}(E)=(\widetilde{U}f )(E)=\int_0^{\infty}dr f(r) 
                             \overline{\chi (r;E)} 
      \label{rhoU}
\end{eqnarray}
that brings ${\cal D}(H)$ onto the space
\begin{equation}
      {\cal D}(\widetilde{E})=\{ \widetilde{f}(E) \in 
                             L^2( (0,\infty ),\rho (E)dE) \, | \  
                             \int_0^{\infty}dE \,
                              E^2|\widetilde{f}(E)|^2 \rho (E)
                             <\infty \} \, .
       \label{rhospace}
\end{equation}
Eqs.~(\ref{rhoU}) and (\ref{rhospace}) provide a $\rho$-diagonalization of 
$H$.  If we seek a $\delta$-diagonalization, i.e., if we seek eigenfunctions
that are $\delta$-normalized, then the measure $\rho (E)$ must be 
absorbed by the eigenfunctions and by the wave functions.\footnote{The 
meaning of the $\delta$-normalization of the eigenfunctions will be explained
in Section~2.9.}  This is why we define 
\begin{equation}
      \sigma (r;E):=\sqrt{\rho (E)} \, \chi (r;E) \, ,
      \label{dnes}
\end{equation}
which is the eigensolution of the differential operator $h$ that 
is $\delta$-normalized, and
\begin{equation}
       \widehat{f}(E):=\sqrt{\rho (E)}{\widetilde f}(E) \, , \quad
       \widetilde{f}(E) \in L^2( (0,\infty ),\rho (E)dE) \, ,
\end{equation}
and construct the unitary operator
\begin{eqnarray}
      \widehat{U}:L^2((0,\infty)),\rho (E)dE) &\longmapsto & L^2((0,\infty),dE)
                                                           \nonumber \\
                        {\widetilde f} &\longmapsto & 
                        \widehat{f}(E)= \widehat{U}{\widetilde f}(E):=
                         \sqrt{\rho (E)}{\widetilde f}(E) \, . 
\end{eqnarray}
The operator that $\delta$-diagonalizes our Hamiltonian is
$U:={\widehat U}{\widetilde U}$,
\begin{eqnarray}
      U:L^2([0,\infty)),dr) &\longmapsto & L^2((0,\infty),dE) \nonumber \\
                       f &\longmapsto & Uf:={\widehat f} \, . 
\end{eqnarray} 
The action of $U$ can be written as an integral operator, 
\begin{equation}
      \widehat{f}(E)=(Uf)(E)=
       \int_0^{\infty}dr f(r) \overline{\sigma (r;E)} \, , \quad
      f(r)\in L^2([0,\infty ),dr) \, .
      \label{inteexpre}
\end{equation}
The image of ${\cal D}(H)$ under the action of $U$ is
\begin{equation}
      {\cal D}(\widehat{E}):=U{\cal D}(H)=
      \{ {\widehat f}(E) \in L^2((0,\infty ),dE) \, | \ \int_0^{\infty}
                         E^2|\widehat{f}(E)|^2 dE<\infty  \} \, .
\end{equation}

Therefore, we have constructed a unitary operator 
\begin{eqnarray}
      U:{\cal D}(H) \subset L^2([0,\infty ),dr) &\longmapsto & 
      {\cal D}(\widehat{E})
       \subset L^2((0,\infty ),dE) \nonumber \\
       f &\longmapsto & \widehat{f}=Uf
\end{eqnarray}
that transforms from the position representation into the energy 
representation.  The operator $U$ diagonalizes our Hamiltonian in the sense 
that $\widehat{E}\equiv UHU^{-1}$ is the multiplication operator,
\begin{eqnarray}
      \widehat{E}:{\cal D}(\widehat{E}) \subset L^2((0,\infty ),dE)  
        &\longmapsto & L^2((0,\infty ),dE)  \nonumber \\
        \widehat{f} & \longmapsto & (\widehat{E} \widehat{f})(E):=
               E\widehat{f}(E) \, .   
\end{eqnarray}
The inverse operator of $U$ is given by (see Theorem~3 of 
Appendix~\ref{sec:A3})
\begin{equation}
      f(r)=(U^{-1}\widehat{f})(r)= 
        \int_0^{\infty}dE\, \widehat{f}(E)\sigma(r,E) \, ,
       \quad \widehat{f}(E)\in L^2((0,\infty ),dE) \, .
     \label{invdiagonaliza}
\end{equation}
The operator $U^{-1}$ transforms from the energy representation into the 
position representation.

The expressions (\ref{inteexpre}) and (\ref{invdiagonaliza}) provide 
the eigenfunction expansion of any square integrable function in terms of the 
eigensolutions $\sigma (r;E)$ of $h$.

The unitary operator $U$ can be looked at as a sort of generalized Fourier 
transform:  the Fourier transform connects the position and the momentum 
representations.  $U$ connects the position and the energy 
representations.  The role played by the plane waves $e^{-ipx}$ (which are 
generalized eigenfunctions of the operator $-id/dx$) is here played by the 
$\sigma (r;E)$ (which are generalized eigenfunctions of the differential 
operator $h$).  Therefore $\sigma (r;E)\equiv \langle r|E\rangle$, which are 
the $\delta$-normalized eigensolutions of the Schr\"odinger equation, can be 
viewed as ``transition elements'' between the $r$- and the 
$E$-representations.

The label $f$ of the functions in the position representation is 
different to the label $\widehat f$ of the functions in the
energy representation, because they have different functional 
dependences.  Similar considerations apply to the Hamiltonian, the domain,
the resolvent, etc.  This
is not the standard practice in the physics literature,
where different representations are usually identified and labeled by the 
same symbol (see, for instance, \cite{NEWTON,COHEN,TAYLOR,BOHM}).

\def\thesubsection{\thesection.\arabic{subsection}}
\subsection{The Need of the Rigged Hilbert Space}
\label{sec:tneodRHS}

The Sturm-Liouville theory only provides a domain ${\cal D}(H)$ on which 
the Hamiltonian $H$ is self-adjoint and a unitary operator $U$ that 
diagonalizes $H$.  This unitary operator induces a direct integral 
decomposition of the Hilbert space 
(see \cite{VON2,GELFAND}),
\begin{eqnarray}
      {\cal H} &\longmapsto & 
      U{\cal H} \equiv \widehat{\cal H}=\oplus \int_{{\rm Sp}(H)}{\cal H}(E)dE 
             \nonumber \\
       f &\longmapsto & Uf\equiv  \{ \widehat{f}(E) \}, \, \quad 
        \widehat{f}(E) \in {\cal H}(E) \, ,
       \label{dirintdec}
\end{eqnarray}
where ${\cal H}$ is realized by $L^2([0,\infty ),dr)$, and $\widehat{\cal H}$ 
is realized by $L^2([0,\infty ),dE)$.  The Hilbert space ${\cal H}(E)$ 
associated to each energy eigenvalue of ${\rm Sp}(H)$ is realized by the
Hilbert space of complex numbers $\mathbb C$.  On $\widehat{\cal H}$, the 
operator $H$ acts as the multiplication operator,
\begin{equation}
       Hf \longmapsto UHf\equiv \{ E\widehat{f}(E) \} 
         \, , \quad f\in {\cal D}(H) \, . 
\end{equation}
The scalar product on $\widehat{\cal H}$ can be written as
\begin{equation}
       \left( \widehat{f},\widehat{g} \right) _{\widehat{\cal H}}=
       \int_{ {\rm Sp}(H)} 
       \left( \widehat{f}(E), \widehat{g}(E) \right) _E dE \, ,
\end{equation}
where the scalar product $(\, \cdot \, , \, \cdot \, )_E$ on 
${\cal H}(E)$ is the usual scalar product on $\mathbb C$,
\begin{equation} 
       \left( \widehat{f}(E), \widehat{g}(E) \right) _E=
      \overline{\widehat{f}(E)} \; \widehat{g}(E) \, . 
\end{equation}

As we shall explain below, the direct integral decomposition does not 
accommodate some of the basic requirements needed in Quantum Mechanics.  These
requirements can be accommodated by the RHS.   

One of the most important principles of Quantum Mechanics is that the quantity
$(\varphi ,H\varphi)$ should fit the experimental expectation value of the 
observable $H$ in the state $\varphi$.  However, $(\varphi ,H\varphi)$ is not 
defined for every element in $\cal H$, but only for those square normalizable 
wave functions that are also in ${\cal D}(H)$.  Therefore, not every square 
normalizable function can represent a ``physical wave function'', but only 
those that are (at least) in ${\cal D}(H)$.  Another fundamental assumption 
of quantum physics is that the quantity 
\begin{equation}
      {\rm disp}_{\varphi}H=(\varphi , H^2\varphi )-(\varphi ,H\varphi )^2 
       \label{dr}
\end{equation}
represents the dispersion of the observable $H$ in the state $\varphi$, and
that
\begin{equation}
       \Delta _{\varphi}H\equiv \sqrt{{\rm disp}_{\varphi}H}
        \label{uncer}
\end{equation}
represents the uncertainty of the observable $H$ in the state 
$\varphi$.  The quantities (\ref{dr}) and (\ref{uncer}) are not defined for
every element of the Hilbert space either.  Therefore, we would like to find a 
subdomain $\mathbf \Phi$ included in ${\cal D}(H)$ on which the expectation
values
\begin{equation}
      (\varphi ,H^n\varphi )\, , \quad n=0,1,2,\ldots\, ,  \quad \varphi \in
      \mathbf \Phi
       \label{ev}
\end{equation}
are well-defined.

Another important requirement of Quantum Mechanics is that algebraic 
operations such as the sum and multiplication of two operators are 
well-defined.  In the HS formalism, these algebraic operations are not always
well-defined because the domains on which these operators are self-adjoint
do not remain stable under their actions in general.  In fact,
much of the trouble of the HS formalism comes from domain questions.  In our
case, the domain ${\cal D}(H)$ in (\ref{domain}) does not remain stable 
under $H$.  We therefore would like to find a subdomain $\mathbf \Phi$ 
included in ${\cal D}(H)$ that remains stable under the action of $H$ and 
all of its powers,
\begin{equation}
      H^n:{\mathbf \Phi} \longmapsto {\mathbf \Phi} \, , \quad n=0,1,2,\ldots
      \label{stabeilis}
\end{equation}
One can see that if Eq.~(\ref{stabeilis}) holds, then the expectation values
(\ref{ev}) are well-defined for each $\varphi$ in $\mathbf \Phi$, i.e., if
the domain $\mathbf \Phi$ remains stable under the action of $H$, then the 
expectation values of $H$ in any state $\varphi \in \mathbf \Phi$ are 
well-defined.

In Quantum Mechanics, it is always assumed that for each $E\in {\rm Sp}(H)$ 
there is a Dirac ket $|E\rangle$ such that 
\begin{equation}
       H^{\times}|E\rangle =E|E\rangle 
       \label{ke}
\end{equation}
and such that the Dirac basis vector expansion~(\ref{DbveI}) 
holds.  Equation~(\ref{ke}) has no solution in 
the Hilbert space when $E$ belongs to the continuous part of the spectrum 
of the Hamiltonian.  In fact, Eq.~(\ref{ke}) has to be related to the 
equation
\begin{equation}
       \langle \vec{x}|H^{\times}|E\rangle =E\langle \vec{x}|E\rangle \, ,
\end{equation}
which in the radial representation reads
\begin{equation}
       h \sigma (r;E)=E\sigma (r;E) \, ,
\end{equation}
where $h$ is the differential operator (\ref{doh}) and $\sigma (r;E)$ is 
the delta-normalized eigenfunction (\ref{dnes}).  Since 
$\sigma (r;E)\equiv \langle r|E\rangle$ lies outside $L^2([0,\infty ),dr)$,
i.e.,
\begin{equation}
       \int_0^{\infty}dr \, |\sigma (r;E)|^2 =\infty \, ,
\end{equation}
the corresponding eigenket $|E\rangle$, which is defined by
\begin{eqnarray}
       |E\rangle :\mathbf \Phi & \longmapsto & \mathbb C \nonumber \\
       \varphi & \longmapsto & \langle \varphi |E\rangle := 
       \int_0^{\infty}\overline{\varphi (r)}\sigma (r;E)dr\, ,
       \label{needefinitionket}
\end{eqnarray}
should also lie outside the Hilbert space.  We shall show that $|E\rangle$ 
is an element of $\mathbf \Phi ^{\times}$.  

In summary, what our mathematical framework should provide us with is:
\begin{enumerate}
      \item a dense invariant domain $\mathbf \Phi$ on which all the powers 
        of $H$ and all the expectation values (\ref{ev}) are well-defined,
      \item smooth enough wave functions so that Eq.~(\ref{ke}) holds in
           the sense
          \begin{equation}
                 \langle \varphi |H^{\times}|E\rangle =
                 E\langle \varphi |E\rangle \, ,
          \end{equation}
      \item any wave function can be expanded by a Dirac basis vector 
        expansion.
\end{enumerate}
In the direct integral decomposition formalism, there is not enough 
room for either of these three requirements.  This is why we introduce the RHS.

\def\thesubsection{\thesection.\arabic{subsection}}
\subsection{Construction of the Rigged Hilbert Space}
\label{sec:RotD}

The first step is to make all the powers of the Hamiltonian 
well-defined.  In order to do so, we construct the maximal invariant subspace 
$\cal D$ of the operator $H$,
\begin{equation}
      {\cal D}:= \bigcap _{n=0}^{\infty}{\cal D}(H^n) \, .
      \label{misus}
\end{equation}
The space $\cal D$ is the largest subspace of ${\cal D}(H)$ that remains 
stable under the action of the Hamiltonian $H$ and all of its powers.  It 
is easy to check that
\begin{eqnarray}
      {\cal D}=\{ \varphi \in L^2([0,\infty ),dr) \, | && \
       h^n\varphi (r)\in L^2([0,\infty ),dr),\
       h^n\varphi (0)=0,  \varphi ^{(n)}(a)=\varphi ^{(n)}(b)=0, \nonumber \\
      && 
       n=0,1,2,\ldots ; \ \varphi (r) \in C^{\infty}([0,\infty)) \} \, .
      \label{mainisexi}
\end{eqnarray}
The conditions $\varphi ^{(n)}(a)=\varphi ^{(n)}(b)=0$ in (\ref{mainisexi}) 
come from taking the discontinuities of the potential $V(r)$ at $r=a$ and at 
$r=b$ into consideration (cf.~\cite{ROBERTS}). 

The second step is to find a subspace $\mathbf \Phi$ on which
the eigenkets $|E\rangle$ of $H$ are well-defined as antilinear 
functionals.  For each $E\in {\rm Sp}(H)$, we associate a ket $|E\rangle$ to 
the generalized eigenfunction $\sigma (r;E)$ through
\begin{eqnarray}
       |E\rangle :\mathbf \Phi & \longmapsto & \mathbb C \nonumber \\
       \varphi & \longmapsto & \langle \varphi |E\rangle := 
       \int_0^{\infty}\overline{\varphi (r)}\sigma (r;E)dr 
       =\overline{(U\varphi )(E)} \, .
       \label{definitionket}
\end{eqnarray}
As actual computations show, the ket $|E\rangle$ in (\ref{definitionket})
is a generalized eigenfunctional of $H$ if $\mathbf \Phi$ is included in the 
maximal invariant subspace of $H$,
\begin{equation}
       \mathbf \Phi \subset  {\cal D} \, .
\end{equation}
Due to the non-square integrability of the eigenfunction $\sigma (r;E)$, we
need to impose further restrictions on the elements of $\cal D$ in order
to make the eigenfunctional $|E\rangle$ in Eq.~(\ref{definitionket})
continuous,
\begin{equation}
      \int_0^{\infty}dr \, \left| (r+1)^n(h+1)^m\varphi (r)\right| ^2<\infty, 
        \quad n,m=0,1,2,\ldots
       \label{condcodogis}
\end{equation}
The imposition of conditions (\ref{condcodogis}) upon the space $\cal D$
leads to the space of wave functions of the square barrier potential,
\begin{equation}
       {\mathbf \Phi} =\{ \varphi \in {\cal D} \, | \ 
       \int_0^{\infty}dr \, \left| (r+1)^n(h+1)^m\varphi (r)\right| ^2<\infty, 
        \quad n,m=0,1,2,\ldots \} \, .
\end{equation}    

On $\mathbf \Phi$, we define the family of norms
\begin{equation}
      \| \varphi \| _{n,m} := 
    \sqrt{\int_0^{\infty}dr \, \left| (r+1)^n(h+1)^m\varphi (r)\right| ^2}
    \, , \quad n,m=0,1,2,\ldots 
      \label{nmnorms}
\end{equation}
The quantities (\ref{nmnorms}) fulfill the conditions to be a norm 
(cf.~Proposition~1 of Appendix~\ref{sec:A4}) and can be used to define a 
countably normed topology $\tau _{\mathbf \Phi}$ on $\mathbf \Phi$ 
(see~\cite{GELFAND}),  
\begin{equation}
      \varphi _{\alpha}\, \mapupdown{\tau_{\mathbf \Phi}}{\alpha \to \infty}
      \, \varphi \quad {\rm iff} \quad  \| \varphi _{\alpha}-\varphi \| _{n,m} 
      \, \mapupdown{}{\alpha \to \infty}\, 0 \, , \quad n,m=0,1,2, \ldots 
\end{equation}       
One can see that the space 
$\mathbf \Phi$ is stable under the action of $H$ and that $H$ is 
$\tau _{\mathbf \Phi}$-continuous (cf.~Proposition~2 of 
Appendix~\ref{sec:A4}). 

Once we have constructed the space $\mathbf \Phi$, we can construct its 
topological dual $\mathbf \Phi ^{\times}$ as the space of 
$\tau _{\mathbf \Phi}$-continuous antilinear functionals on 
$\mathbf \Phi$ (see~\cite{GELFAND}) and therewith the RHS of the 
square barrier potential (for $l=0$) 
\begin{equation}
       {\mathbf \Phi} \subset L^2([0,\infty ),dr) \subset 
        {\mathbf \Phi}^{\times}  \, .
\end{equation}

The ket $|E\rangle$ in Eq.~(\ref{definitionket}) is a 
well-defined antilinear functional on $\mathbf \Phi$, i.e., $|E\rangle$ 
belongs to $\mathbf \Phi ^{\times}$  (cf.~Proposition~3 of 
Appendix~\ref{sec:A4}).  The ket $|E\rangle$ is a generalized eigenvector of 
the Hamiltonian $H$ (cf.~Proposition~3 of Appendix~\ref{sec:A4}),
\begin{equation}
       H^{\times}|E\rangle=E|E\rangle \, ,
\end{equation}
i.e.,
\begin{equation}
       \langle \varphi |H^{\times}|E\rangle=
        \langle H\varphi |E\rangle = E\langle \varphi|E\rangle \, ,
       \quad \forall \varphi \in \mathbf \Phi \, .
       \label{afegenphis}
\end{equation}        

On the space $\mathbf \Phi$, all the expectation values of the Hamiltonian and 
all the algebraic operations involving $H$ are well-defined, and the 
generalized eigenvalue equation (\ref{afegenphis}) holds.  As we shall see in 
the next section, the functions $\varphi$ of $\mathbf \Phi$ can be 
expanded by a Dirac basis vector expansion.

\def\thesubsection{\thesection.\arabic{subsection}}
\subsection{Dirac Basis Vector Expansion}
\label{sec:DBVE}

We are now in a position to derive the Dirac basis vector expansion.  This 
derivation consists of the restriction of the Weyl-Kodaira expansions
(\ref{inteexpre}) and (\ref{invdiagonaliza}) to the space 
$\mathbf \Phi$.  If we denote $\langle r|\varphi \rangle \equiv \varphi (r)$ 
and $\langle E|r\rangle \equiv \overline{\sigma (r;E)}$, and if we define the 
action of the left ket $\langle E|$ on $\varphi \in \mathbf \Phi$ as 
$\langle E| \varphi \rangle := \widehat{\varphi}(E)$, then 
Eq.~(\ref{inteexpre}) becomes
\begin{equation}
      \langle E|\varphi \rangle =\int_0^{\infty}dr \, 
                                  \langle E |r \rangle 
                                  \langle r|\varphi \rangle \, , \quad
       \varphi \in \mathbf \Phi \, .
        \label{DFeps}
\end{equation}
If we denote $\langle r|E \rangle \equiv \sigma (r;E)$, then 
Eq.~(\ref{invdiagonaliza}) becomes
\begin{equation}
      \langle r|\varphi \rangle =\int_0^{\infty}dE \,
      \langle r|E \rangle  \langle E|\varphi \rangle \, , \quad
       \varphi \in \mathbf \Phi \, .
       \label{inveqDva}
\end{equation}
This equation is the Dirac basis vector expansion of the square barrier 
potential.  In fact, when we formally write (\ref{DbveI}) in the 
position representation, we get to (\ref{inveqDva}).

In Eq.~(\ref{inveqDva}), the wave function 
$\langle r|\varphi \rangle$ is spanned in a ``Fourier-type'' expansion by 
the eigenfunctions $\langle r|E \rangle$.  In this expansion, each 
eigenfunction $\langle r|E \rangle$ is weighted by 
$\langle E|\varphi \rangle =\widehat{\varphi}(E)$, which is
the value of the wave function in the energy representation at the 
point $E$.  Thus any function $\varphi (r)=\langle r|\varphi \rangle$
of $\mathbf \Phi$ can be written as a linear superposition of the
monoenergetic eigenfunctions $\sigma (r;E)=\langle r|E \rangle$.

Although the Weyl-Kodaira expansions (\ref{inteexpre}) and 
(\ref{invdiagonaliza}) are valid for every element of the Hilbert space,
the Dirac basis vector expansions (\ref{DFeps}) and (\ref{inveqDva}) are only 
valid for functions $\varphi \in  \mathbf \Phi$ because only those 
functions fulfill both
\begin{equation}
      \overline{\widehat{\varphi}(E)}=\langle \varphi |E\rangle
\end{equation}
and
\begin{equation}
       \langle \varphi |H^{\times}|E\rangle = \langle H\varphi |E\rangle =
       E \langle \varphi|E\rangle \, .
\end{equation}

Another way to rephrase the Dirac basis vector expansion is the Nuclear 
Spectral (Gelfand-Maurin) theorem.  Instead of using the general statement of 
\cite{GELFAND}, we prove this theorem using the 
Sturm-Liouville theory (see Proposition~4 of Appendix~\ref{sec:A5}).  The
Nuclear Spectral Theorem allows us to write the scalar product of any 
two functions $\varphi ,\psi$ of $\mathbf \Phi$ in terms of the action of
the kets $|E\rangle$ on $\varphi ,\psi$:
\begin{equation}
      (\varphi ,\psi )=\int_0^{\infty} dE \,
      \langle \varphi |E\rangle \langle E|\psi \rangle \, , \quad 
       \forall \varphi ,\psi \in \mathbf \Phi \, .
       \label{GMT1a}
     \end{equation}
It also allows us to write the matrix elements of the Hamiltonian and all
of its powers between two elements $\varphi ,\psi$ of $\mathbf \Phi$
in terms of the action of the kets $|E\rangle$ on $\varphi ,\psi$:
\begin{equation}
      (\varphi ,H^n \psi )=\int_0^{\infty}dE \,
      E^n \langle \varphi |E\rangle \langle E|\psi \rangle \, , \quad 
       \forall \varphi ,\psi \in {\mathbf \Phi} \, , n=1,2,\ldots
        \label{GMT2a}
\end{equation}

\def\thesubsection{\thesection.\arabic{subsection}}
\subsection{Energy Representation of the RHS}
\label{ESrepr}

In this section, we construct the energy representation of the RHS.  Since
the unitary operator $U$ transforms from the position representation into the 
energy representation, the action of $U$ on the RHS provides the energy 
representation of the RHS. 

We have already shown that in the energy representation the Hamiltonian $H$
acts as the multiplication operator $\widehat{E}$.  The energy representation
of the space $\mathbf \Phi$ is defined as  
\begin{equation}
      \widehat{\mathbf \Phi}:= U\mathbf \Phi \, .
\end{equation}
It is very easy to see that $\widehat{\mathbf \Phi}$ is a linear subspace
of $L^2([0,\infty ),dE)$.  In order to endow $\widehat{\mathbf \Phi}$ with a 
topology $\tau _{\widehat{\mathbf \Phi}}$, we carry the topology on 
$\mathbf \Phi$ into $\widehat{\mathbf \Phi}$,
\begin{equation}
      \tau _{\widehat{\mathbf \Phi}}:=U\tau _{\mathbf \Phi} \, .
\end{equation}
With this topology, the space $\widehat{\mathbf \Phi}$ is a linear topological
space.  If we denote the dual space of $\widehat{\mathbf \Phi}$ by
$\widehat{\mathbf \Phi}^{\times}$, then we have
\begin{equation}
      U^{\times}{\mathbf \Phi}^{\times}=
      (U{\mathbf \Phi})^{\times}= \widehat{\mathbf \Phi}^{\times} \, .
\end{equation}
If we denote $|\widehat{E}\rangle \equiv U^{\times}|E\rangle$, then we can 
prove that $|\widehat{E}\rangle$ is the antilinear Schwartz delta functional, 
i.e., $|\widehat{E}\rangle$ is the antilinear functional that associates to 
each function $\widehat{\varphi}$ the complex conjugate of its value at the 
point $E$ (see Proposition~5 of Appendix~\ref{sec:A6}),
\begin{eqnarray}
      |\widehat{E}\rangle: \widehat{\mathbf \Phi} &\longmapsto & 
       \mathbb C \nonumber \\
       \widehat{\varphi} &\longmapsto & 
        \langle \widehat{\varphi}|\widehat{E}\rangle :=
        \overline{ \widehat{\varphi}(E)} \, .
\end{eqnarray}
Therefore, the Schwartz delta functional appears in the (spectral) energy 
representation of the RHS associated to the Hamiltonian.  If we write the
action of the Schwartz delta functional as an integral operator, then the
Dirac $\delta$-function appears as the kernel of that integral operator.  

It is very helpful to show the different realizations of the RHS through the
following diagram:
\begin{equation}
      \begin{array}{ccclccclccll}
      H; & \varphi (r) & \ & \mathbf \Phi & \subset & L^2([0,\infty),dr) &
      \subset & \mathbf \Phi ^{\times} & \  & |E\rangle & \ &
      {\rm position \ repr.} \nonumber \\  [2ex]
       & & \ & \downarrow U &  &\downarrow U  &
       & \downarrow U^{\times} & \ & & &  \nonumber \\ [2ex]  
      \widehat{E}; & \widehat{\varphi}(E) & \ & \widehat{\mathbf \Phi} & 
       \subset & 
      L^2([0,\infty),dE) & \subset & \widehat{\mathbf \Phi} ^{\times} & 
      \ & |\widehat{E}\rangle &\ & {\rm energy \ repr.}  \\ 
      \end{array}
      \label{diagramsavp}
\end{equation}
On the top line of the diagram (\ref{diagramsavp}), we have the RHS, the 
Hamiltonian, the wave functions and the Dirac kets in the position 
representation.  On the bottom line, we have their energy representation 
counterparts.

\def\thesubsection{\thesection.\arabic{subsection}}
\subsection{Meaning of the $\delta$-Normalization of the Eigenfunctions}
\label{sec:MFM}

In this section, we show that the $\delta$-normalization of the eigenfunctions 
is related to the measure $d\mu (E)$ that is used to compute the scalar 
product of the wave functions in the energy representation,
\begin{equation}
      (\varphi , \psi )=\int_0^{\infty}\overline{\varphi (E)} \psi (E) 
        d\mu (E) \, .
      \label{sprusduyms}
\end{equation}
We will see that if the measure in (\ref{sprusduyms}) is the Lebesgue measure 
$dE$, then the eigenfunctions are $\delta$-normalized, and that if the 
measure is $\rho (E)dE$, then the eigenfunctions are 
$\rho$-normalized.

For the sake of simplicity, in this section we label the wave functions in 
the position and in the energy representation with the same symbol.  With
this notation, Eq.~(\ref{inveqDva}) reads
\begin{mathletters}
      \label{expnsofisn}
\begin{eqnarray}
       &&\varphi (r)=\int_0^{\infty}dE\, \varphi (E) \sigma (r;E) \, , \\
       && \psi (r)= \int_0^{\infty}dE\, \psi (E) \sigma (r;E) \, .
\end{eqnarray}
\end{mathletters}
Since $\varphi (r), \psi (r)\in L^2( [0,\infty),dr)$, their scalar product
is well-defined,
\begin{equation}
       (\varphi ,\psi )=\int_0^{\infty}dr \, \overline{\varphi (r)}\psi (r)\, .
       \label{scparlpodof}
\end{equation}
Plugging (\ref{expnsofisn}) into (\ref{scparlpodof}), we obtain
\begin{equation} 
      (\varphi ,\psi )=\int_0^{\infty}dE\int_0^{\infty}dE'\, 
        \overline{\varphi (E)} \psi (E')\int_0^{\infty}dr \, 
        \overline{\sigma (r;E)}\sigma (r;E') \, .
       \label{fe1}
\end{equation}
If we use the Lebesgue measure $dE$, then the scalar product (\ref{sprusduyms})
can be written as
\begin{equation}
       (\varphi ,\psi )=\int_0^{\infty}dE\, \overline{\varphi (E)}\psi (E) \, .
        \label{form2}
\end{equation}
Comparison of (\ref{fe1}) and (\ref{form2}) leads to 
\begin{equation}
      \int_0^{\infty}dr \, \overline{\sigma (r;E)}\sigma (r;E')=
       \delta (E-E') \, ,
       \label{formdeno}
\end{equation}
i.e., the eigenfunctions $\sigma (r;E)$ are $\delta$-normalized.  

We now consider the case in which the eigenfunctions are 
$\rho$-normalized.  If we use the measure $d\mu (E)=\rho (E) dE$, then  
the scalar product of $\varphi$ and $\psi$ is given by 
\begin{equation}
     (\varphi ,\psi )=\int_0^{\infty} 
      \overline{\varphi _{\rho}(E)}\psi _{\rho}(E) \rho (E)\, dE \, ,
      \label{rhospsiner}
\end{equation}
where $\varphi _{\rho}(E):=\varphi (E)/\sqrt{\rho (E)}$ and 
$\psi _{\rho}(E):= \psi (E)/\sqrt{\rho (E)}$.  If we define
$\sigma _{\rho}(r;E):=\sigma (r;E)/\sqrt{\rho (E)}$, then 
Eq.~(\ref{expnsofisn}) reads 
\begin{mathletters}
      \label{rhoexpnsofisn}
\begin{eqnarray}
       &&\varphi (r)=\int_0^{\infty}\varphi _{\rho}(E) \sigma _{\rho}(r;E)
         \rho (E) \, dE \, , \\
       && \psi (r)= \int_0^{\infty}\psi _{\rho}(E) \sigma _{\rho}(r;E)
          \rho (E) \, dE  \, .
\end{eqnarray}
\end{mathletters}
Plugging Eq.~(\ref{rhoexpnsofisn}) into (\ref{scparlpodof}), we obtain
\begin{equation}
      (\varphi ,\psi )=\int_0^{\infty}dE\int_0^{\infty}dE'\, 
        \overline{\varphi _{\rho}(E)} \psi _{\rho}(E')\rho (E) \rho (E')
        \int_0^{\infty}dr \, 
        \overline{\sigma _{\rho}(r;E)}\sigma _{\rho}(r;E') \, .
      \label{compsexps}
\end{equation}
Comparison of (\ref{compsexps}) and (\ref{rhospsiner}) leads to
\begin{equation}
      \int_0^{\infty}dr \, \overline{\sigma _{\rho}(r;E)}\sigma _{\rho}(r;E')=
       \frac{1}{\rho (E)}\delta (E-E') \, ,
\end{equation}
i.e., the eigenfunctions $\sigma _{\rho}(r;E)$ are $\rho$-normalized.

\def\thesection{\arabic{section}}
\section{Conclusion}
\def\thesection{\arabic{section}}
\setcounter{equation}{0}
\label{sec:C}

In this paper, we have constructed the Rigged Hilbert Space
\begin{equation}
      {\mathbf \Phi} \subset L^2([0,\infty),dr) \subset {\mathbf \Phi}^{\times}
       \label{ctriplet}
\end{equation}
of the square barrier Hamiltonian and its energy representation
\begin{equation}
       \widehat{\mathbf \Phi} \subset L^2([0,\infty),dE) \subset 
        \widehat{\mathbf \Phi} ^{\times} \, .
\end{equation}
The spectrum of the Hamiltonian $H$ is the positive real semiaxis.  For each 
value $E$ of the spectrum
of $H$, we have constructed a Dirac ket $|E\rangle$ that is a generalized 
eigenfunctional of $H$ whose corresponding generalized eigenvalue is 
$E$.  In the energy representation, $|E\rangle$ acts as
the antilinear Schwartz delta functional.  On the space $\mathbf \Phi$, all
algebraic operations involving the Hamiltonian $H$ are well-defined.  In 
particular, the expectation values of the Hamiltonian in any element of 
$\mathbf \Phi$ are well-defined.  Any element of 
$\mathbf \Phi$ can be expanded in terms of the eigenkets $|E\rangle$ by a 
Dirac basis vector expansion.  The elements of
$\mathbf \Phi$ are represented by well-behaved functions, in contrast to 
the elements of the Hilbert space, which are represented by sets of 
equivalent functions that can vary arbitrarily on any set of zero
Lebesgue measure.  Therefore, it seems natural to conclude that a physically 
acceptable wave function is not any element of the Hilbert space but rather 
an element of the subspace $\mathbf \Phi$.

The monoenergetic eigensolutions $\sigma (r;E)$ of the time-independent
Schr\"odinger equation are not square integrable, and therefore they cannot
represent an acceptable wave function.  Those eigensolutions has been used to
define the eigenkets $|E\rangle$ that expand the physical wave
functions $\varphi \in \mathbf \Phi$ in a Dirac basis vector expansion.  The
eigenkets $|E\rangle$ belong to the space $\mathbf \Phi ^{\times}$.  The
eigensolutions $\sigma (r;E)=\langle r|E\rangle$ have been also used
to construct the unitary operator $U$ that transforms from the position
representation into the energy representation.

In our quest for the RHS of the square barrier potential, we have found a 
systematic method to construct the RHS of a large class of spherically 
symmetric potentials:
\begin{enumerate}
      \item Expression of the formal differential operator.
      \item Hilbert space $\cal H$ of square integrable functions on which the 
            formal differential operator acts.
      \item A domain ${\cal D}(H)$ of the Hilbert space on which the formal 
            differential operator is self-adjoint.
      \item Green functions (resolvent) of that self-adjoint operator.
      \item Diagonalization of the self-adjoint operator, eigenfunction 
            expansion of the elements of $\cal H$ in terms of the 
            eigensolutions of the formal differential operator, and direct 
            integral decomposition of $\cal H$ induced by
            the self-adjoint operator.
      \item Subspace $\mathbf \Phi$ of ${\cal D}(H)$ on which all the 
            expectation values of $H$ are well-defined and on which the Dirac
            kets act as antilinear functionals.
      \item Rigged Hilbert space 
            $\mathbf \Phi \subset {\cal H}\subset \mathbf \Phi ^{\times}$. 
\end{enumerate}

\def\thesection{\arabic{section}}
\section*{Acknowledgments}
\setcounter{equation}{0}

One of the authors (R.~de la Madrid) wishes to thank Prof.~J.-P.~Antoine and 
T.~Kuna for introducing him to the Sturm-Liouville theory and 
Profs.~R.~De la Llave, A.~Galindo, L.~Caffarelli and J.~Bona for 
illuminating discussions.  Many thanks are due to N.~Harshman and C.~Koeninger
for invaluable advise on English style.  

Financial support from the U.E.~TMR Contract number ERBFMRX-CT96-0087 ``The 
Physics of Quantum Information'', from the Welch Foundation, from La Junta 
de Castilla y Le\'on Project PC02 1999, from DGCYT PB98-0370, and from
DGICYT PB98-0370 is gratefully acknowledged.

\appendix
\def\thesection{\Alph{section}}
\section{Self-Adjoint Extension}
\setcounter{equation}{0}
\label{sec:A1}

In this appendix, we list the possible self-adjoint extensions associated to 
the differential operator $h$.  We first need some definitions 
(cf.~\cite{DUNFORD}).

\vskip0.5cm

{\bf Definition~1}\quad By $AC^2 ([0,\infty ) )$ we denote the space of 
all functions $f$ which have a continuous derivative in 
$[0,\infty )$, and for which $f'$ is not only continuous but also 
absolutely continuous over each compact subinterval of $[0,\infty )$.  Thus
$f ^{(2)}$ exists almost everywhere, and is integrable over any compact 
subinterval of $[0,\infty )$.  At $0$ $f'$ is continuous from the right.

\vskip0.5cm

The space $AC ^2 ([0,\infty ) )$ is the largest space of functions on which 
the differential operator $h$ can be defined. 

\vskip0.5cm

{\bf Definition~2}\quad  We define the spaces
\begin{eqnarray}
      &&{\cal H}^2 _h ([0,\infty ) ):= \{ f\in AC ^2 ([0,\infty )) \, | \   
                                       f, hf \in L^2 ( [0,\infty ),dr) \} \\
      &&{\cal H}^2  ([0,\infty ) ):= \{ f\in AC ^2 ([0,\infty ) \, | \ f,
                                       f^{(2)} \in L^2 ( [0,\infty ),dr) \} \\
      &&{\cal H}^2 _0 ([0,\infty ) ):= \{ f\in {\cal H}^2([0,\infty )) \, | \ 
         f \  \mbox{vanishes outside some compact subset of} \ (0,\infty ) \}
        \, .
\end{eqnarray}

Using these spaces, we can define the necessary operators to calculate the 
self-adjoint extensions associated to $h$.

\vskip1cm

{\bf Definition~3}\quad  If $h$ is the formal differential operator 
(\ref{doh}), we define the operators $H_0$ and $H_1$ on 
$L^2 ([0,\infty ),dr )$ by the formulas
\begin{eqnarray}
      &&{\cal D}(H_0)={\cal H}^2 _0 ([0,\infty ) ), \quad H_0f:=hf, \quad
       f \in {\cal D}(H_0) \, . \\
      &&{\cal D}(H_1)={\cal H}^2 _h ([0,\infty ) ), \quad H_1f:=hf, \quad
       f\in {\cal D}(H_1) \, .     
\end{eqnarray} 

\vskip0.5cm

The operators $H_0$ and $H_1$ are sometimes called the {\it minimal} 
and the {\it maximal} operators associated to the differential operator $h$, 
respectively.  The domain ${\cal D}(H_1)$ is the largest domain of
the Hilbert space $L^2 ([0,\infty ),dr )$ on which the action of
the differential operator $h$ can be defined and remains inside 
$L^2 ([0,\infty ),dr )$.  Further, $H_0^{\dagger}=H_1$.

The self-adjoint extensions of $H_0$ are given by the restrictions
of the operator $H_1$ to domains determined by the conditions (see
\cite{DUNFORD}, page 1306)
\begin{equation}
       f(0)+\alpha  \, f'(0)=0 \, ,
       \quad -\infty < \alpha \leq \infty \, .
\end{equation}
These boundary conditions lead to the domains
\begin{equation}
      {\cal D}_{\alpha}(H)=\{ f \in {\cal D}(H_1) \, | \ 
      f(0)+\alpha  \, f'(0)=0 \} \, , \quad
       -\infty < \alpha \leq \infty \, .
\end{equation} 
On these domains, the formal differential operator $h$ is self-adjoint.  The 
boundary condition that fits spherically symmetric potentials is $f(0)=0$, 
i.e., $\alpha =0$.  This condition selects our domain (\ref{domain}),
\begin{equation}
      {\cal D}(H)={\cal D}_{\alpha =0}(H) =
      \{ f \in {\cal D}(H_1) \, | \  f(0)=0 \} \, .
\end{equation}

\def\thesection{\Alph{section}}
\section{Resolvent and Green Function}
\setcounter{equation}{0}
\label{sec:A2}

The following theorem provides the procedure to compute the Green 
function of the Hamiltonian $H$ (cf.~Theorem~XIII.3.16 of 
Ref.~\cite{DUNFORD}):

\vskip0.5cm

{\bf Theorem~1}\quad  Let $H$ be the self-adjoint operator (\ref{operator}) 
derived from the real formal differential operator (\ref{doh}) by the 
imposition of the boundary condition (\ref{sac}).  Let $\Im E \neq 0$.  Then 
there is exactly one solution $\chi (r;E)$ of $(h-E)\sigma =0$ 
square-integrable at $0$ and satisfying the boundary condition (\ref{sac}), 
and exactly one solution $\Theta (r;E)$ of $(h-E)\sigma =0$ square-integrable 
at infinity.  The resolvent $(E-H)^{-1}$ is an integral operator whose kernel 
$G(r,s;E)$ is given by
\begin{equation}
       G(r,s;E)=\left\{ \begin{array}{ll}
               \frac{2m}{\hbar ^2} \,
      \frac{\chi (r;E) \, \Theta (s;E)}{W(\chi ,\Theta )}
               &r<s \\ 
      \frac{2m}{\hbar ^2} \,
      \frac{\chi (s;E) \, \Theta (r;E)}{W(\chi ,\Theta )}
                       &r>s  \, ,
                  \end{array} 
                 \right. 
	\label{exofGFA}
\end{equation}
where $W(\chi ,\Theta )$ is the Wronskian of $\chi$ and $\Theta$
\begin{equation}
       W(\chi ,\Theta )=\chi \Theta '-\chi ' \Theta \, .
\end{equation}

\vskip0.5cm

If we define 
\begin{mathletters}
\begin{eqnarray}
       &&\widetilde{k} :=\sqrt{-\frac{2m}{\hbar ^2}\, E} \, , \\
       && \widetilde{Q}:=\sqrt{-\frac{2m}{\hbar ^2}\, (E-V_0)} \, ,
\end{eqnarray}
\end{mathletters}
then the functions $\widetilde{{\cal J}}(E)$ of 
Eq.~(\ref{tildechifunction}) are given by
\begin{mathletters}
      \label{tildejfunc}
\begin{eqnarray}
      &&\widetilde{{\cal J}}_1(E)=
       \frac{1}{2}e^{-\widetilde{Q}a}
             \left[ 
     \left(1+\frac{\widetilde{k}}
      {\widetilde{Q}}\right) 
     e^{\widetilde{k}a} +
     \left(-1+\frac{\widetilde{k}}
     {\widetilde{Q}}
     \right)
     e^{-\widetilde{k}a}
     \right] , \\
     && \widetilde{{\cal J}}_2(E)=
        \frac{1}{2}e^{\widetilde{Q}a}
             \left[
       \left(1-\frac{\widetilde{k}}
            {\widetilde{Q}} \right)
       e^{\widetilde{k}a} +
        \left(-1-\frac{\widetilde{k}}
            {\widetilde{Q}} \right) 
         e^{-\widetilde{k}a}
     \right] , \\
     && \widetilde{{\cal J}}_3(E)=\frac{1}{2}e^{-\widetilde{k}b}
                   \left[
      \left(1+\frac{\widetilde{Q}}
      {\widetilde{k}}\right) 
      e^{\widetilde{Q}b}\widetilde{{\cal J}}_1(E)+
     \left(1-\frac{\widetilde{Q}}
            {\widetilde{k}} \right) 
      e^{-\widetilde{Q}b}\widetilde{{\cal J}}_2(E)
     \right]  , \\ 
     &&\widetilde{{\cal J}}_4(E)=\frac{1}{2}e^{\widetilde{k}b}
                   \left[
     \left(1-\frac{\widetilde{Q}}
            {\widetilde{k}} \right) 
     e^{\widetilde{Q}b}\widetilde{{\cal J}}_1(E)+
     \left(1+\frac{\widetilde{Q}}
            {\widetilde{k}} \right) 
      e^{-\widetilde{Q}b}\widetilde{{\cal J}}_2(E)
     \right]  ,  
\end{eqnarray}
\end{mathletters}
and the functions $\widetilde{\cal A}(E)$ of Eq.~(\ref{tildethetfunc}) by
\begin{mathletters}
      \label{tildeAfunc}
\begin{eqnarray}
      && \widetilde{{\cal A}}_3(E)=
      \frac{1}{2}e^{-\widetilde{Q}b} 
     \left(1-\frac{\widetilde{k}}
      {\widetilde{Q}}\right) 
     e^{-\widetilde{k}b} , \\
     && \widetilde{{\cal A}}_4(E)=
       \frac{1}{2}e^{\widetilde{Q}b}
       \left(1+\frac{\widetilde{k}}
            {\widetilde{Q}} \right)
       e^{-\widetilde{k}b} , \\
    && \widetilde{{\cal A}}_1(E)=\frac{1}{2}e^{-\widetilde{k}a}
                   \left[
      \left(1+\frac{\widetilde{Q}}
      {\widetilde{k}}\right) 
      e^{\widetilde{Q}a} \widetilde{{\cal A}}_3(E)+
     \left(1-\frac{\widetilde{Q}}
            {\widetilde{k}} \right) 
      e^{-\widetilde{Q}a}  \widetilde{{\cal A}}_4(E)
     \right] , \\ 
     && \widetilde{ {\cal A}}_2(E)=
     \frac{1}{2}e^{\widetilde{k}a}
                   \left[
     \left(1-\frac{\widetilde{Q}}
            {\widetilde{k}} \right) 
     e^{\widetilde{Q}a} \widetilde{{\cal A}}_3(E)+
     \left(1+\frac{\widetilde{Q}}
            {\widetilde{k}} \right) 
      e^{-\sqrt{- \widetilde{Q}}a} \widetilde{{\cal A}}_4(E)
     \right] . 
\end{eqnarray}
\end{mathletters}
The expression for the Wronskian of 
$\widetilde{\chi}$ and $\widetilde{\Theta}_-$ is
\begin{equation}
      W(\widetilde{\chi} ,\widetilde{\Theta}_- )=
      -2\widetilde{k} \widetilde{\cal{J}}_3(E) \, .
\end{equation}

If we define
\begin{mathletters}
\begin{eqnarray}
      &&k:=\sqrt{\frac{2m}{\hbar ^2} \, E} \, , \\
      &&Q:=\sqrt{\frac{2m}{\hbar ^2} \, (E-V_0)} \, ,
\end{eqnarray}
\end{mathletters}
then the functions ${\cal J}(E)$ of Eq.~(\ref{chi}) are given
by
\begin{mathletters}
      \label{Jfunction}
\begin{eqnarray}
      &&{\cal J}_1(E)=\frac{1}{2}e^{-iQa} 
     \left( \sin (ka)+
      \frac{k}{iQ}
      \cos (ka) \right) , \\
      &&{\cal J} _2 (E)=\frac{1}{2}e^{iQa} 
     \left( \sin (ka)-
      \frac{k}{iQ}
      \cos (ka) \right) , \\
      &&{\cal J}_3(E)=\frac{1}{2}e^{-ikb}
                   \left[
            \left(1+ \frac{Q}{k} \right) 
       e^{iQb}{\cal J}_1(E)+
       \left(1-\frac{Q}{k} \right) 
       e^{-iQb}{\cal J}_2(E) 
        \right] , \\
       &&{\cal J}_4(E)=\frac{1}{2}e^{ikb}
                   \left[
            \left(1-\frac{Q}{k} \right) 
       e^{iQb}{\cal J} _1(E)+
       \left(1+\frac{Q}{k}\right) 
        e^{-iQb}{\cal J}_2(E) 
        \right] , 
\end{eqnarray}
\end{mathletters}
and the functions ${\cal A}^+(E)$ of Eq.~(\ref{theta+fun}) by
\begin{mathletters}
      \label{A+functions}
\begin{eqnarray}
     &&{\cal A}_3^+(E)=\frac{1}{2}e^{-iQb} 
     \left(1+\frac{k}{Q}\right) e^{ikb} , \\
     &&{\cal A}_4^+(E)=\frac{1}{2}e^{iQb}
       \left(1-\frac{k}{Q} \right)
       e^{ikb} , \\
     &&{\cal A}_1^+(E)=\frac{1}{2}e^{-ika}
                   \left[
      \left(1+\frac{Q}{k}\right) 
      e^{iQa}{\cal A}_3^+(E)+
     \left(1-\frac{Q}{k} \right) 
      e^{-iQa} {\cal A}_4^+(E)
     \right] , \\ 
     &&{\cal A}_2^+(E)=\frac{1}{2}e^{ika}
                   \left[
     \left(1-\frac{Q}{k} \right) 
     e^{iQa}{\cal A}_3^+(E)+
     \left(1+\frac{Q}{k} \right) 
      e^{-iQa}{\cal A}_4^+(E)
     \right] . 
\end{eqnarray}
\end{mathletters}
The Wronskian of $\chi$ and $\Theta _+$ is
\begin{equation}
       W(\chi , \Theta _+)=2ik {\cal J}_4(E) \, .
\end{equation}
The functions ${\cal A}^-(E)$ of Eq.~(\ref{thetafun-}) are given by
\begin{mathletters}
       \label{A-functions}
\begin{eqnarray}
    &&{\cal A}_3^-(E)=\frac{1}{2}e^{-iQb} 
     \left(1-\frac{k}{Q}\right) 
     e^{-ikb} , \\
    &&{\cal A}_4^-(E)=\frac{1}{2}e^{iQb}
       \left(1+\frac{k}{Q} \right)
       e^{-ikb} , \\
    &&{\cal A}_1^-(E)=\frac{1}{2}e^{-ika}
                   \left[
      \left(1+\frac{Q}{k}\right) 
      e^{iQa}{\cal A}_3^-(E)+
     \left(1-\frac{Q}{k} \right) 
      e^{-iQa}{\cal A}_4^-(E)
     \right] , \\ 
    &&{\cal A}_2^-(E)=\frac{1}{2}e^{ika}
                   \left[
     \left(1-\frac{Q}{k} \right)e^{iQa}{\cal A}_3^-(E) +
     \left(1+\frac{Q}{k} \right) 
      e^{-iQa}{\cal A}_4^-(E)
     \right]  . 
\end{eqnarray}
\end{mathletters}
The Wronskian of $\chi$ and $\Theta _-$ is
\begin{equation}
       W(\chi , \Theta _-)=-2ik {\cal J}_3(E) \, .
\end{equation}

\def\thesection{\Alph{section}}
\section{Diagonalization and Eigenfunction \newline Expansion}
\setcounter{equation}{0}
\label{sec:A3}

The theorem that provides the operator $U$ that diagonalizes $H$ is 
(cf.~Theorem XIII.5.13 of Ref.~\cite{DUNFORD})

\vskip0.5cm

{\bf Theorem~2} (Weyl-Kodaira)\quad  Let $h$ be the formally self-adjoint 
differential operator (\ref{doh}) defined on the interval 
$[0,\infty )$.  Let $H$ be the self-adjoint operator (\ref{operator}).  Let
$\Lambda$ be an open interval of the real axis, and suppose that there 
is given a set $\{ \sigma _1(r;E),\, \sigma _2(r;E)\}$ of functions, defined 
and continuous on $(0,\infty )\times \Lambda$, such that for each fixed 
$E$ in $\Lambda$, $\{ \sigma _1(r;E),\, \sigma _2(r;E)\}$ forms a basis for
the space of solutions of $h\sigma =E\sigma$.  Then there exists a 
positive $2\times 2$ matrix measure $\{ \rho _{ij} \}$ defined on
$\Lambda$, such that
\begin{enumerate}
      \item the limit 
     \begin{equation}
      (U f)_i(E)=\lim_{c\to 0}\lim_{d\to \infty} 
        \left[ \int_c^d f(r) \overline{\sigma _i(r;E)}dr \right]
      \end{equation}
     exists in the topology of $L^2(\Lambda ,\{ \rho _{ij}\})$ for each 
    $f$ in $L^2([0,\infty ),dr)$ and defines an isometric isomorphism $U$ of
    $E(\Lambda )L^2([0,\infty ),dr)$ onto $L^2(\Lambda ,\{ \rho _{ij}\})$;
      \item for each Borel function $G$ defined on the real line and vanishing
       outside $\Lambda$,
    \begin{equation}
      U{\cal D}(G(H))=\{ [f_i]\in L^2(\Lambda ,\{ \rho _{ij}\}) \, | \
           [Gf_i]\in L^2(\Lambda ,\{ \rho _{ij}\}) \}
     \end{equation}
    and
    \begin{equation}
      (UG(H)f)_i(E)=G(E)(Uf)_i(E), \quad i=1,2, \, E\in \Lambda ,
       \, f\in {\cal D}(G(H)) \, .
    \end{equation}  
\end{enumerate}

\vskip0.5cm

The theorem that provides the inverse of the operator $U$ is 
(cf.~Theorem XIII.5.14 of Ref.~\cite{DUNFORD})

\vskip0.5cm

{\bf Theorem~3} (Weyl-Kodaira)\quad  Let $H$, $\Lambda$, 
$\{ \rho _{ij} \}$, etc., be as in Theorem~2.  Let $E_0$ and $E_1$ be the end 
points of $\Lambda$.  Then
\begin{enumerate}
      \item the inverse of the isometric isomorphism $U$ of 
      $E(\Lambda )L^2([0,\infty ),dr)$ onto $L^2(\Lambda ,\{ \rho _{ij}\})$ is 
      given by the formula
     \begin{equation}
       (U^{-1}F)(r)=\lim_{\mu _0 \to E_0}\lim_{\mu _1 \to E_1}
       \int_{\mu _0}^{\mu _1} \left( \sum_{i,j=1}^{2}
             F_i(E)\sigma _j(r;E)\rho _{ij}(dE) \right)
     \end{equation}
    where $F=[F_1,F_2]\in L^2(\Lambda ,\{ \rho _{ij}\})$, the limit existing
   in the topology of $L^2([0, \infty ),dr)$;
    \item if $G$ is a bounded Borel function vanishing outside a Borel set 
    $e$ whose closure is compact and contained in $\Lambda$, then $G(H)$ has 
    the representation
   \begin{equation}
       G(H)f(r)=\int _0^{\infty}f(s)K(H,r,s)ds \, , 
   \end{equation}
   where
   \begin{equation}
      K(H,r,s)=\sum_{i,j=1}^2 \int_e 
       G(E)\overline{\sigma _i(s;E)}\sigma _j(r;E)\rho _{ij}(dE) \, .
   \end{equation}
\end{enumerate}

\vskip0.5cm

The spectral measures are provided by the following theorem 
(cf.~Theorem XIII.5.18 of Ref.~\cite{DUNFORD}):

\vskip0.5cm

{\bf Theorem~4} (Titchmarsh-Kodaira)\quad  Let $\Lambda$ be an open interval 
of the real axis and $O$ be an open set in the complex plane containing 
$\Lambda$.  Let $\{ \sigma _1(r;E),\, \sigma _2(r;E)\}$ be a set of functions 
which form a basis for the solutions of the equation $h\sigma =E\sigma$, 
$E\in O$, and which are continuous on $(0,\infty )\times O$ and analytically 
dependent on $E$ for $E$ in $O$.  Suppose that the kernel $G(r,s;E)$ for the 
resolvent $(E-H)^{-1}$ has a representation
\begin{equation}
      G(r,s;E)=\left\{ \begin{array}{lll}
                   \sum_{i,j=1}^2 \theta _{ij}^-(E)\sigma _i(r;E)
                   \overline{\sigma _j(s;\overline{E})}\, , &
                     \qquad & r<s \, ,  \\
                  \sum_{i,j=1}^2 \theta _{ij}^+(E)\sigma _i(r;E)
                   \overline{\sigma _j(s;\overline{E})} \, ,&\qquad & r>s \, ,
                  \end{array}
                 \right.
\end{equation}
for all $E$ in ${\rm Re}(H)\cap O$, and that $\{ \rho _{ij} \}$ is a 
positive matrix measure on $\Lambda$ associated with $H$ as in Theorem 2.  Then
the functions $\theta _{ij}^{\pm}$ are analytic in ${\rm Re}(H)\cap O$, and
given any bounded open interval $(E_1,E_2)\subset \Lambda$, we have for 
$1\leq i,j\leq 2$,
\begin{equation}
       \begin{array}{lll}
       \rho _{ij}((E_1,E_2))&=& \lim_{\delta \to 0}\lim_{\epsilon \to 0+}
         \frac{1}{2\pi i}\int_{E_1+\delta}^{E_2-\delta}
          [ \theta _{ij}^-(E-i\epsilon )-\theta _{ij}^-(E+i\epsilon )
          ]dE \\ 
      \quad &=& \lim_{\delta \to 0}\lim_{\epsilon \to 0+}
         \frac{1}{2\pi i}\int_{E_1+\delta}^{E_2-\delta}
          [ \theta _{ij}^+(E-i\epsilon )-\theta _{ij}^+(E+i\epsilon )
          ]dE \, .
         \end{array} 
\end{equation}

\vskip0.5cm

The functions $\widetilde{\cal B}(E)$ of Eq.~(\ref{tildesigma1}) are given by
\begin{mathletters}
      \label{tildeBfunctions}
\begin{eqnarray}
      && \widetilde{{\cal B}}_3(E)=
      \frac{1}{2}e^{-\widetilde{Q}b} 
     \left(1+\frac{\widetilde{k}}{\widetilde{Q}}\right)
     e^{\widetilde{k}b} , \\
     && \widetilde{{\cal B}}_4(E)=
       \frac{1}{2}e^{\widetilde{Q}b}
       \left(1-\frac{\widetilde{k}}{\widetilde{Q}} \right)
       e^{\widetilde{k}b}  , \\
    && \widetilde{{\cal B}}_1(E)=\frac{1}{2}e^{-\widetilde{k}a}
                   \left[
      \left(1+\frac{\widetilde{Q}}{\widetilde{k}}\right) 
      e^{\widetilde{Q}a} \widetilde{{\cal B}}_3(E)+
     \left(1-\frac{\widetilde{Q}}{\widetilde{k}} \right) 
      e^{-\widetilde{Q}a}  \widetilde{{\cal B}}_4(E)
     \right] , \\ 
     && \widetilde{ {\cal B}}_2(E)=
     \frac{1}{2}e^{\widetilde{k}a}
                   \left[
     \left(1-\frac{\widetilde{Q}}{\widetilde{k}} \right) 
     e^{\widetilde{Q}a} \widetilde{{\cal B}}_3(E)+
     \left(1+\frac{\widetilde{Q}}{\widetilde{k}} \right) 
      e^{-\widetilde{Q}a} \widetilde{{\cal B}}_4(E)
     \right] .
\end{eqnarray}
\end{mathletters}
The functions ${\cal C}(E)$ of Eq.~(\ref{sigam2cos}) are given by
\begin{mathletters}
      \label{Cfunctions}
\begin{eqnarray}
      &&{\cal C}_1(E)= 
       \frac{1}{2}e^{-iQa}
             \left( \cos (ka)-\frac{k}{iQ} \sin (ka) \right), \\
     && {\cal C}_2(E)=
       \frac{1}{2}e^{iQa}
        \left( \cos (ka)+\frac{k}{iQ} \sin (ka) \right), \\
     && {\cal C}_3(E)=\frac{1}{2}e^{-ikb}
                   \left[
      \left(1+\frac{Q}{k}\right) e^{iQb}{\cal C}_1(E)+
     \left(1-\frac{Q}{k} \right) e^{-iQb}{\cal C}_2(E)
     \right] , \\ 
     &&{\cal C}_4(E)=\frac{1}{2}e^{ikb}
                   \left[
     \left(1-\frac{Q}{k} \right) e^{iQb}{\cal C}_1(E)+
     \left(1+\frac{Q}{k} \right) 
      e^{-iQb}{\cal C}_2(E)
     \right]  . 
\end{eqnarray}
\end{mathletters}

\def\thesection{\Alph{section}}
\section{Construction of the RHS}
\setcounter{equation}{0}
\label{sec:A4}

{\bf Proposition~1} \quad  The quantities
\begin{equation}
    \| \varphi \| _{n,m} := 
    \sqrt{\int_0^{\infty}dr \, \left| (r+1)^n(h+1)^m\varphi (r)\right| ^2}, 
     \quad \varphi \in {\mathbf \Phi} \, , \,  n,m=0,1,2,\ldots, 
      \label{anmnorms}
\end{equation}
are norms.

\vskip0.2cm

{\it Proof}\quad  It is very easy to show that the quantities (\ref{anmnorms}) 
fulfill the conditions to be a norm,
\begin{mathletters}
\begin{eqnarray}
      &&\| \varphi +\psi \| _{n,m} \leq \| \varphi \| _{n,m} + 
       \| \psi \| _{n,m} \, , \\
      && \| \alpha \varphi \| _{n,m}=|\alpha |\, \| \varphi \| _{n,m} \, , \\
      && \| \varphi \| _{n,m} \geq 0 \, , \\
      && {\rm If }\  \| \varphi \| _{n,m} =0, \ {\rm then} \ \varphi =0 \, .
        \label{homiensi}
\end{eqnarray}
\end{mathletters}
The only condition that is somewhat difficult to prove is (\ref{homiensi}): if 
$\| \varphi \| _{n,m}=0$, then 
\begin{equation}
       (1+r)^n(h+1)^m\varphi (r)=0 \, ,
\end{equation}
which yields
\begin{equation}
      (h+1)^m\varphi (r)=0 \, .  
      \label{homiodhiis}
\end{equation}
If $m=0$, then Eq.~(\ref{homiodhiis}) implies $\varphi (r)=0$.  If $m=1$, then 
Eq.~(\ref{homiodhiis}) implies that $-1$ is an eigenvalue of $H$ whose 
corresponding eigenvector is $\varphi$.  Since $-1$ is not an
eigenvalue of $H$, $\varphi$ must be the zero vector.  If $m>1$, the proof 
is similar.

\vskip0.5cm

{\bf Proposition~2}\quad  The space $\mathbf \Phi$ is stable under the action 
of $H$, and $H$ is $\tau _{\mathbf \Phi}$-continuous.  

\vskip0.2cm

{\it Proof}\quad  In order to see that $H$ is 
$\tau _{\mathbf \Phi}$-continuous, we just have to realize that
\begin{eqnarray}
      \| H\varphi \| _{n,m}&=&\| (H+I)\varphi -\varphi \| _{n,m} \nonumber \\
       &\leq & \| (H+I)\varphi \| _{n,m}+ \| \varphi \| _{n,m} \nonumber \\
       &=&\| \varphi \| _{n,m+1}+\| \varphi \| _{n,m} \, .
       \label{tauphiscont}
\end{eqnarray}
We now prove that $\mathbf \Phi$ is stable under the action of $H$.  Let
$\varphi \in \mathbf \Phi$.  To say that $\varphi \in \mathbf \Phi$ is
equivalent to say that $\varphi \in {\cal D}$ and
that the norms $\| \varphi \| _{n,m}$ are finite for every 
$n,m=0,1,2,\ldots \ $  Since $H\varphi$ is also in ${\cal D}$, and since the 
norms $\| H\varphi \| _{n,m}$ are also finite (see Eq.~(\ref{tauphiscont})), 
the vector $H\varphi$ is also in $\mathbf \Phi$.

\vskip0.5cm

{\bf Proposition~3}\quad The function
\begin{eqnarray}
       |E\rangle :\mathbf \Phi & \longmapsto & \mathbb C \nonumber \\
       \varphi & \longmapsto & \langle \varphi |E\rangle := 
       \int_0^{\infty}\overline{\varphi (r)}\sigma (r;E)dr 
       =\overline{(U\varphi )(E)} \, .
       \label{adefinitionket}
\end{eqnarray}
is an antilinear functional on $\mathbf \Phi$ that is a generalized 
eigenvector of (the restriction to $\mathbf \Phi$ of) $H$.  

\vskip0.2cm

{\it Proof}\quad From the definition (\ref{adefinitionket}), it is pretty easy 
to see that $|E\rangle$ is an antilinear functional.  In order to show that 
$|E\rangle$ is continuous, we define
\begin{equation}
      {\cal M}(E):= \sup _{r\in [0,\infty )} \left| \sigma (r;E) \right| \, .
\end{equation}
Since
\begin{eqnarray}
      |\langle \varphi |E\rangle | &=& |\overline{U\varphi (E)}| \nonumber \\
      &=&\left| \int_0^{\infty}dr \, \overline{\varphi (r)}\sigma(r;E)\right| 
      \nonumber \\
      &\leq & \int_0^{\infty}dr \, |\overline{\varphi (r)}| |\sigma(r;E)|
      \nonumber \\
      &\leq & {\cal M}(E) \int _0^{\infty}dr \, |\varphi (r)| \nonumber \\
      &=& {\cal M}(E) \int_0^{\infty}dr \,
      \frac{1}{1+r} (1+r) |\varphi (r)| \nonumber \\
      &\leq & {\cal M}(E) \left( \int_0^{\infty}dr \, 
      \frac{1}{(1+r)^2} \right) ^{1/2} 
      \left( \int_0^{\infty}dr \, 
      \left| (1+r) \varphi (r) \right| ^2 \right) ^{1/2} \nonumber \\
      &=&{\cal M}(E) \left( \int_0^{\infty}dr \, 
      \frac{1}{(1+r)^2} \right) ^{1/2} 
       \| \varphi \| _{1,0}     \nonumber \\
       &=&{\cal M}(E) \| \varphi \| _{1,0} \, ,
\end{eqnarray}
the functional $|E\rangle$ is continuous when $\mathbf \Phi$ is endowed with 
the $\tau _{\mathbf \Phi}$ topology.

In order to prove that $|E\rangle$ is a generalized eigenvector of $H$, we 
make use of the conditions (\ref{mainisexi}) and (\ref{condcodogis})
satisfied by the elements of $\mathbf \Phi$,
\begin{eqnarray}
       \langle \varphi |H^{\times}|E\rangle &=& \langle H\varphi |E\rangle
       \nonumber \\
       &=& \int_0^{\infty}dr \, 
       \left( -\frac{\hbar ^2}{2m}\frac{d^2}{dr^2}+V(r) \right)
       \overline{\varphi(r)} \sigma (r;E) \nonumber \\
       &=&-\frac{\hbar ^2}{2m}
       \left[ \frac{d\overline{\varphi (r)}}{dr} \sigma(r;E) 
       \right] _0^{\infty} 
       +\frac{\hbar ^2}{2m}
       \left[ \overline{\varphi (r)} \frac{d\sigma(r;E)}{dr} 
       \right] _0^{\infty} \nonumber \\ 
       &&+ \int_0^{\infty}dr \, \overline{\varphi(r)}
       \left( -\frac{\hbar ^2}{2m}\frac{d^2}{dr^2}+V(r) \right)\sigma (r;E)
        \nonumber \\
       &=&E\langle \varphi |E\rangle \, .
\end{eqnarray}

Similarly, one can also prove that
\begin{equation}
       \langle \varphi | (H^{\times})^n|E\rangle =
       E^n \langle \varphi |E\rangle \, .
\end{equation}

\def\thesection{\Alph{section}}
\section{Dirac Basis Vector Expansion}
\setcounter{equation}{0}
\label{sec:A5}

{\bf Proposition~4} (Nuclear Spectral Theorem) \quad Let 
\begin{equation}
      {\mathbf \Phi} \subset L^2([0,\infty ),dr)\subset \mathbf \Phi ^{\times}
\end{equation}
be the RHS of the square barrier Hamiltonian $H$ such that $\mathbf \Phi$ 
remains invariant under $H$ and $H$ is a $\tau _{\mathbf \Phi}$-continuous 
operator on $\mathbf \Phi$.  Then, for each $E$ in 
the spectrum of $H$ there is a generalized eigenvector $|E\rangle$ such that
\begin{equation}
       H^{\times}|E\rangle =E|E\rangle
\end{equation}
and such that
\begin{equation}
      (\varphi ,\psi )=\int_{ {\rm Sp}(H)}dE\, 
      \langle \varphi |E\rangle \langle E|\psi \rangle \, , \quad 
       \forall \varphi ,\psi \in \mathbf \Phi \, ,
       \label{GMT1}
\end{equation}
and
\begin{equation}
      (\varphi ,H^n \psi )=\int_{ {\rm Sp}(H)}dE \,
      E^n \langle \varphi |E\rangle \langle E|\psi \rangle \, , \quad 
       \forall \varphi ,\psi \in {\mathbf \Phi} \, , n=1,2,\ldots
        \label{GMT2}
\end{equation}

\vskip0.2cm

{\it Proof} \quad Let $\varphi$ and $\psi$ be in $\mathbf \Phi$.  Since $U$ 
is unitary,
\begin{equation}
       (\varphi ,\psi )=(U\varphi ,U\psi )=(\widehat{\varphi} ,\widehat{\psi})
       \, .
       \label{Usiuni}
\end{equation}
The wave functions $\widehat{\varphi}$ and $\widehat{\psi}$ are in particular
elements of $L^2([0,\infty ),dE)$.  Therefore their scalar product is 
well-defined,
\begin{equation}
      (\widehat{\varphi} ,\widehat{\psi} )=
      \int_{{\rm Sp}(H)}dE \, 
      \overline{ \widehat{\varphi}(E)} \widehat{\psi}(E) \, .
      \label{sphatvhaps}
\end{equation}
Since $\varphi$ and $\psi$ belong to $\mathbf \Phi$, the action of each 
eigenket $|E\rangle$ on them is well-defined,
\begin{mathletters}
         \label{actionofEpsi}
\begin{eqnarray}
      \langle \varphi |E\rangle =\overline{ \widehat{\varphi}(E)} \, ,\\
      \langle E|\psi \rangle =\widehat{\psi}(E) \, .
\end{eqnarray}
\end{mathletters}
Plugging Eq.~(\ref{actionofEpsi}) into Eq.~(\ref{sphatvhaps}) and
Eq.~(\ref{sphatvhaps}) into Eq.~(\ref{Usiuni}), we get to 
Eq.~(\ref{GMT1}).  The proof of (\ref{GMT2}) is similar:
\begin{eqnarray}
       (\varphi ,H^n\psi )&=&(U\varphi , UH^nU^{-1}U\psi ) \nonumber \\
       &=& (\widehat{\phi} ,\widehat{E}^n\widehat{\psi} ) \nonumber \\
       &=&\int_{{\rm Sp}(H)}dE \,\overline{ \widehat{\varphi}(E)} 
        (\widehat{E}^n\widehat{\psi})(E) \nonumber \\
       &=&\int_{{\rm Sp}(H)}dE\, E^n \overline{ \widehat{\varphi}(E)} 
        \widehat{\psi}(E) \nonumber \\
       &=& \int_{{\rm Sp}(H)}dE \,
      E^n \langle \varphi |E\rangle \langle E|\psi \rangle  \, . 
\end{eqnarray}

\def\thesection{\Alph{section}}
\section{Energy Representation of the RHS}
\setcounter{equation}{0}
\label{sec:A6}

{\bf Proposition~5} \quad The energy representation of the eigenket 
$|E\rangle$ is the antilinear Schwartz delta functional $|\widehat{E}\rangle$.

\vskip0.2cm

{\it Proof} \quad Since 
\begin{eqnarray}
       \langle \widehat{\varphi }|U^{\times}|E\rangle &=&
       \langle U^{-1}\widehat{\varphi }|E\rangle  \nonumber \\
       &=& \langle \varphi |E\rangle \nonumber \\
       &=& \int_0^{\infty}\overline{\varphi (r)}\sigma (r;E)dr \nonumber \\
       &=& \overline{ \widehat{\varphi}(E)} \, ,
\end{eqnarray}
the functional $U^{\times}|E\rangle =|\widehat{E}\rangle$ is the antilinear
Schwartz delta functional.


\begin{thebibliography}{99}

\bibitem{DIRAC} P.~A.~M.~Dirac, {\it The principles of Quantum Mechanics},
3rd ed., Clarendon Press, Oxford (1947).

\bibitem{VON} J.~von Neumann, {\it Mathematische Grundlagen der 
Quantentheorie}, Springer, Berlin (1931); English translation by R.~T.~Beyer, 
Princeton University Press, Princeton (1955).

\bibitem{SCHWARTZ} L.~Schwartz, {\it Th\'eory de Distributions}, Hermann,
Paris (1950).

\bibitem{VON2} J.~von Neumann, Ann.~Math.~(N.~Y.) {\bf 50}, 401 (1949). 

\bibitem{GELFAND} I.~M.~Gel'fand, N.~Y.~Vilenkin, 
{\it Generalized Functions, Vol.~4} Academic Press, New York, (1964);
K.~Maurin, {\it Generalized Eigenfunction Expansions and Unitary 
Representations of Topological Groups}, Polish Scientific Publishers,
Warsaw (1968). 

\bibitem{MEJLBO} L.~C.~Mejlbo, Math.~Scand.~{\bf 13}, 129 (1963).

\bibitem{KRISTENSEN} P.~Kristensen, L.~Mejlbo, E.~Thue Poulsen, 
Commun.~Math.~Phys.~{\bf 1} 175 (1965); Math.~Scand.~{\bf 14} 129 (1964).

\bibitem{B60} A.~Bohm, {\it Rigged Hilbert Spaces}, International 
Center for Theoretical Physics Lecture Notes, Publication ICTP 64/9,
Trieste (1964); {\it Boulder Lectures in Theoretical Physics, 1966},
Volume 9A, Gordon and Breach, New York (1967).

\bibitem{ROBERTS} J.~E.~Roberts, J.~Math.~Phys.~{\bf 7}, 1097 (1966);
Commun.~Math.~Phys., {\bf 3}, 98 (1966).

\bibitem{ANTOINE} J.~P.~Antoine, J.~Math.~Phys.~{\bf 10}, 53 (1969);
J.~Math.~Phys.~{\bf 10}, 2276 (1969).

\bibitem{BOHM81} A.~Bohm, J.~Math.~Phys.~{\bf 22}, 2813 (1981).

\bibitem{BOHMGA} A.~Bohm, M.~Gadella, {\it Dirac Kets, Gamow Vectors and
Gelfand Triplets}, Springer Lecture Notes in Physics, {\bf 348}, Berlin (1989).

\bibitem{ANTONIOU93} I.~Antoniou, S.~Tasaki, Int.~J.~Quantum Chemistry
{\bf 46}, 425 (1993).

\bibitem{BOLLINI96} C.~G.~Bollini, O.~Civitarese, A.~L.~De Paoli, M.~C.~Rocca,
J.~Math.~Phys.~{\bf 37}, 4235 (1996). 

\bibitem{BOHM97} A.~Bohm, M.~Loewe, S.~Maxson, P.~Patuleanu, C.~Puntmann and
M.~Gadella, J.~Math.~Phys., {\bf 38}, 6072 (1997).

\bibitem{ANTOINE98} J.-P.~Antoine, in {\it Irreversibility and Causality}, 
edited by A.~Bohm, H.-D.~Doebner, and P.~Kielanowski, Springer-Verlag (1998),
page~3.

\bibitem{DIS} R.~de la Madrid, {\it Quantum Mechanics in Rigged Hilbert Space
Language}, PhD Thesis, Universidad de Valladolid (2001).  Available at
\texttt{http://www.isi.it/$\sim$rafa/}.

\bibitem{BOHM} A.~Bohm, {\it Quantum Mechanics: Foundations and 
Applications}, Springer-Verlag, New York (1986). 

\bibitem{GALINDO} A.~Galindo, P.~Pascual, {\it Mec\'anica Cu\'antica},
Universidad-Manuales, Eudema (1989); English translation by J.~D.~Garc{\'\i}a
and L.~Alvarez-Gaum\'e, Springer-Verlag (1990).

\bibitem{B70} A.~Bohm, {\it The Rigged Hilbert Space and Quantum Mechanics},
Lecture Notes in Physics, vol.~{\bf 78}, Springer, New York (1978). 

\bibitem{DUNFORD} N.~Dunford, J.~Schwartz,
{\it Linear operators}, vol.~II., Interscience Publishers, New York (1963).  

\bibitem{NEWTON} R.~G.~Newton, 
{\it Scattering Theory of Waves and Particles}, McGraw-Hill, New York,
(1966); 2nd edition, Springer-Verlag, New York (1982). 

\bibitem{COHEN} C.~Cohen-Tannoudji, B.~Diu, F.~Lalo\"e, 
{\it Quantum Mechanics, Volumes I and II}, Wiley, New York (1977). 

\bibitem{TAYLOR} J.~R.~Taylor, {\it Scattering theory}, Jhon Wiley \& 
Sons, Inc., New York (1972).

\bibitem{CSF} R.~de la Madrid, Chaos, Solitons \& Fractals {\bf 12},
2689 (2001); {\sf quant-ph/0107096}.

\end{thebibliography}
\end{document}